\documentclass[universe,article,accept,pdftex,moreauthors]{Definitions/mdpi}

\newcommand{\ms}[1]{{#1}}

\firstpage{1} 
\pubvolume{1}
\issuenum{1}
\articlenumber{0}
\pubyear{2023}
\copyrightyear{2023}
\externaleditor{Academic Editor: Firstname Lastname}
\datereceived{ } 
\daterevised{ } 
\dateaccepted{ } 
\datepublished{ } 
\hreflink{https://doi.org/} 

\Title{Alpha Centauri: Disc Dynamics, Planet Stability \& Detectability} 

\TitleCitation{Alpha Centauri: Disc Dynamics, Planet Stability, and Detectability}

\Author{Nicol\'as 
 Cuello $^{1,2,}$*$^{,\dagger}$ 
\orcidA{} and Mario Sucerquia $^{3,4,}$*$^{,\dagger}$\orcidB{}}

\AuthorNames{Nicol\'as Cuello and Mario Sucerquia}

\AuthorCitation{Cuello, N.; Sucerquia, M.}

\address{%
$^{1}$ \quad Univ. Grenoble Alpes, CNRS, IPAG, 38000 Grenoble, France\\
$^{2}$ \quad UFR PHysique, Ing\'enierie, Terre, Environnement, M\'ecanique de l'Universit\'e de Grenoble Alpes, France
\\
$^{3}$ \quad Departamento de Ciencias, Facultad de Artes Liberales, Universidad Adolfo Ibáñez, Av. Padre Hurtado 750, Viña del Mar, Chile 
\\
$^{4}$ \quad N\'ucleo Milenio Formaci\'on Planetaria --- NPF, Universidad de Valpara\'iso, Av. Gran Bretaña 1111, \mbox{Valpara\'iso, Chile}}

\corres{Correspondence: nicolas.cuello@univ-grenoble-alpes.fr; mario.sucerquia@edu.uai.cl}

\firstnote{These authors contributed equally to this work.}

\abstract{Alpha Centauri is a triple stellar system, and it contains the closest star to Earth (Proxima Centauri). Over the last decades, the stars in Alpha Cen and their orbits have been investigated in great detail. However, the possible scenarios for planet formation and evolution in this triple stellar system remain to be explored further. First, we present a 3D hydrodynamical simulation of the circumstellar discs in the binary Alpha Cen~AB. Then, we compute stability maps for the planets within Alpha Cen obtained through N-body integrations. Last, we estimate the radial velocity (RV) signals of such planets. We find that the circumstellar discs within the binary cannot exceed 3~au in radius and that the available dust mass to form planets is about 30~$M_\oplus$. Planets around A and B are stable if their semimajor axes are below 3~au, while those around C are stable and remain unperturbed by the binary AB. For rocky planets, the planetary mass has only a mild effect on the stability. Therefore, Alpha Cen could have formed and hosted rocky planets around each star, which may be detected with RV methods in the future. The exoplanetary hunt in this triple stellar system must continue.}


\begin{document}


\section{Introduction}
\label{sec:intro}

Alpha Centauri stands as the closest known stellar system to our Solar System at~a distance of 1.34~pc away from Earth \citep{Akeson2021}. As~such, it constitutes a target of choice for numerous studies related to stellar and orbit characterisation, e.g., \citep{Pourbaix2016, Liseau2016, Kervella2017}. This system is composed of three stars: Alpha Cen~A, Alpha Cen~B, and~Proxima Centauri (or Alpha Cen~C), which are in a hierarchical orbital configuration. In~other words, the~binary AB semimajor axis is significantly smaller than the semimajor of C around AB \citep{Kervella2017}. The~age of the Alpha Cen system is estimated to be $5.3 \pm 0.3$~Gyr \citep{Thevenin2002, Joyce2018, Akeson2021}, which is similar to our own Solar System, which is approximately 4.6 Gyr old. Recent discoveries have further heightened interest in the Alpha Centauri system. As~a matter of fact, an~Earth-sized exoplanet (called Proxima Centauri~b) was discovered in the habitable zone of Proxima Centauri in 2016 \citep{Anglada2016}. Subsequent research has hinted at the presence of additional planets \citep{Gratton2020, Damasso2020, Faria2022, Wagner2021}, although~confirmation is still~pending.

Regarding the primary components, Alpha Centauri A and B bear a striking resemblance to our Sun in terms of size, mass, and temperature \citep{Kervella2017, Kervella2017b}. Alpha Centauri A is about 10\% more massive ($M_{\rm A}=1.0788\,M_\odot$), whilst Alpha Centauri B is around 90\% as massive as the Sun ($M_{\rm B}=0.9092\,M_\odot$). The~binary orbit is highly eccentric ($0.52$), has a semimajor axis of $23.3$~au, and~has a period of $P_{\rm b} \approx 80$~yr \citep{Akeson2021}. The~values of periastron and apoastron are equal to 11 and 35~au, respectively, which are akin to the mean radial distances of Neptune and Pluto from the Sun. The~third member, Proxima Centauri, is a red dwarf star situated approximately 8200~au away from the central pair AB \citep{Akeson2021}. Thanks to high-precision absolute radial velocity measurements, it is now robustly established that Proxima Centauri is bound to AB and that it takes $511$~kyr to complete a single orbit around the inner \mbox{binary \citep{Kervella2017, Akeson2021}}. Hence, the~closest stars to Earth belong to a triple stellar~system.

Besides its proximity, this system offers an unparalleled laboratory for astrophysical research---including the search for exoplanets \citep{Anglada2016} and the tantalising possibility of extraterrestrial life \citep{Endl2015, Turbet2016}. In~this work, we revisit Alpha Centauri in the context of protoplanetary disc dynamics and planet stability. In~Section~\ref{sec:disc_dyn}, we present an updated hydrodynamical simulation of the circumstellar discs in Alpha Cen using the most precise binary orbit available \citep{Akeson2021}. In~Section~\ref{sec:stability}, we compute stability maps for massless and massive (S-type) planets in Alpha Cen. In~Section~\ref{sec:detectability}, we showcase the radial velocity (RV) signals of the hypothetical (stable) rocky planets in Alpha Cen. Finally, we discuss our results in Section~\ref{sec:discussion}.

\section{Protoplanetary Disc Dynamics in Alpha~Centauri}
\label{sec:disc_dyn}

We ran a 3D hydrodynamical simulation to study the circumstellar discs in a binary system using the {\tt Phantom} smoothed particle hydrodynamics (SPH) code \citep{Price2018}. We modeled the gaseous discs without any dust components for computational costs. Therefore, we assumed that the dust particles are perfectly coupled to the gas and that the total dust mass is around 1\% of the gas mass (as is for typical Class II protoplanetary discs). In~this section, we neglect the gravitational perturbations caused by Alpha Cen~C given the large semimajor axis and that we are primarily interested in the circumstellar discs around stars A and~B.

\subsection{Binary~Setup}
\label{sec:bin_setup}

We initialised the binary orbital parameters using the most recent and accurate orbital solution for Alpha Cen based on highly precise astrometric measurements \citep{Akeson2021}. We set the following values for the semimajor, eccentricity, inclination, argument of periastron, and~longitude of the ascending node: $a_{\rm b}=23.3$~au (corresponding to $17.493$~arcsec), $e_{\rm b}=0.51947$, $\omega_{\rm b}=231.519\deg$, and~$\Omega_{\rm b}=205.073\deg$, respectively. The~true anomaly $f_{\rm b}$ was initially set equal to $180\deg$ to begin with the stars at the apoastron. The~binary was modelled using sink particles~\cite{Bate1995} with both sink radii equal to $0.1$~au. The~masses of the primary (A) and the secondary (B) were set to $M_{\rm A}=1.0788\,M_\odot$ and $M_{\rm B}=0.9092\,M_\odot$, respectively, which gives a mass ratio of $q=M_{\rm B}/M_{\rm A} \approx 0.84$. The~binary orbital parameters are able to evolve during the simulation as the stars accrete gaseous particle from the discs. However, given that the binary angular momentum largely dominates the total angular momentum budget of the system, the~binary orbital parameters remain practically~unchanged.

\subsection{Circumstellar Discs~Setup}
\label{sec:discs_setup}

We followed the disc implementation of {\tt Phantom} and initialised two circumstellar discs for a binary with the same orbital parameters as Alpha Cen AB. Both discs are coplanar with the binary orbital plane. Each disc was modelled using 500,000 SPH particles assuming a total gas mass of $0.01\,M_\odot$ akin to typical protoplanetary disc masses \citep{Miotello2023}. Initially, we set the inner and the outer edges of both discs at $R_{\rm in} = 0.5$~au and $R_{\rm out}=10$~au, respectively. The~surface density profiles followed the same prescription: $\Sigma_{\rm A}(R) = \Sigma_{\rm B}(R) = 2350 \, {\rm g~cm}^{-2} \, \left( R / 1~{\rm au} \right) \left( 1 - \sqrt{0.5\,{\rm au}/R} \right)$, where $R$ is the cylindrical radial coordinate (from the host star) given in au. We adopted a mean Shakura–Sunyaev disk viscosity $\alpha_{\rm SS} \approx 0.005$ by setting a fixed artificial viscosity parameter $\alpha_{\rm AV} = 0.25$ and using the `disk viscosity' flag of {\tt Phantom} \citep{Lodato2010}. We further assumed that the disc is vertically isothermal, and we noted $H$ as the pressure scale height. The~disc aspect ratio $H/R$ at a distance of 1~au was set to $0.05$ for both discs, as~is for typical protoplanetary discs. The~temperature profiles are given by the following: $T_{\rm A}(R) = 690 \, \left(R / 1~{\rm au} \right)^{-0.5}$~K and $T_{\rm B}(R) = 581 \, \left(R / 1~{\rm au} \right)^{-0.5}$~K, thereby considering that stars A and B do not have the same mass and~luminosity.

\subsection{Circumstellar Discs in Alpha~Centauri}
\label{sec:discs_alphacen}

The aim of this new hydrodynamical simulation was to quantify the process of disc truncation in order to assess the mass retention and exchange after repeated stellar interactions. We note that a similar numerical setup was recently proposed for the disc around B in Alpha Cen~\cite{Martin2020} using slightly different values for the binary orbit. Here, the~main difference is that we model both discs simultaneously and consider a revised value of the binary semimajor axis: $a_{\rm b}=23.3$~au \citep{Akeson2021} instead of $23.75$~au. Given the previous exploration of a broad range of disc parameters \citep{Martin2020}, we limit ourselves to a single case. We consider this scenario as an illustrative example of the formation scenario in Alpha~Centauri.

\begin{figure}[H]
\includegraphics[width=\linewidth]{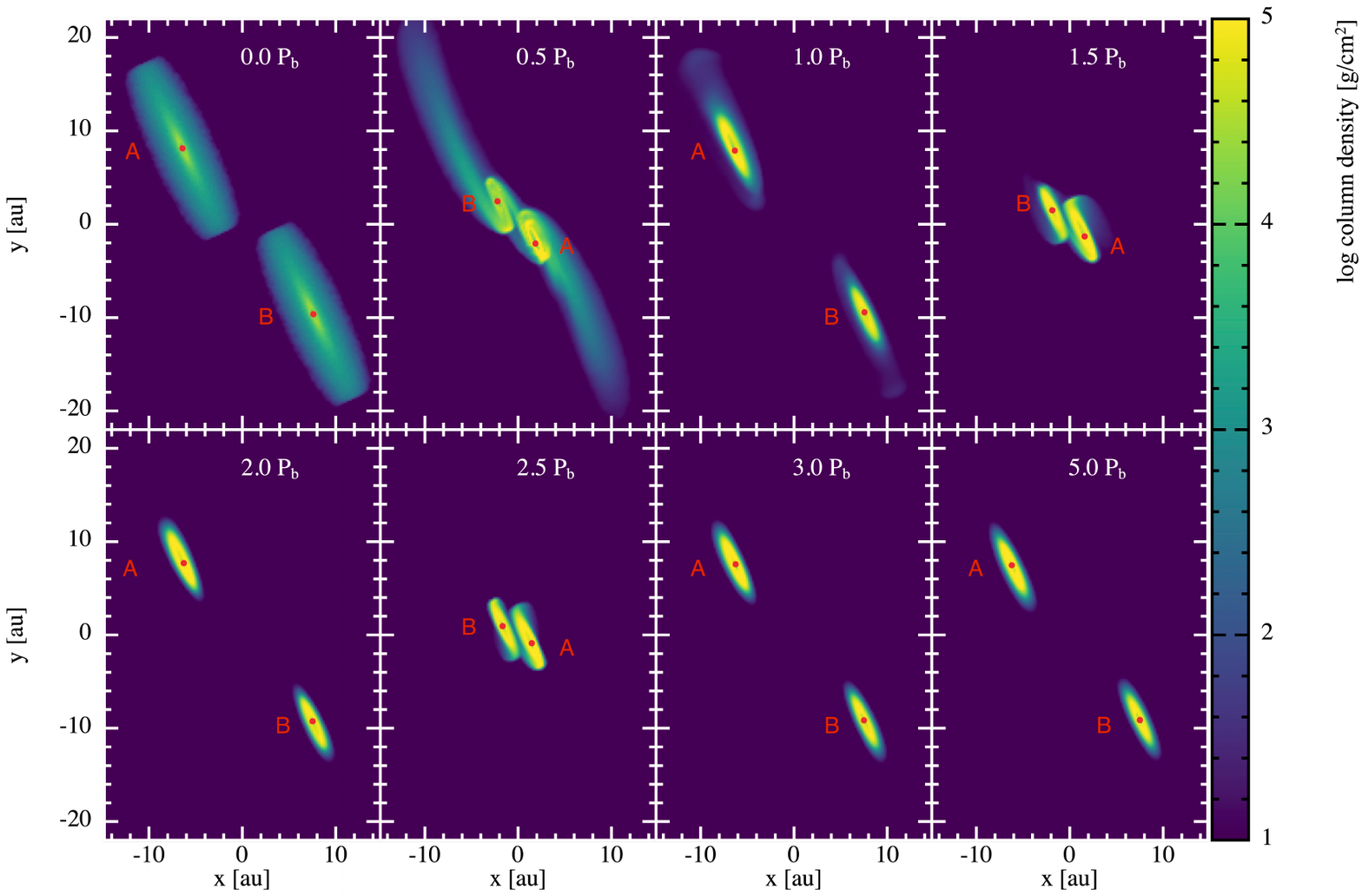}
\caption{Gas column density of the {\tt Phantom} hydrodynamical simulation of the circumstellar discs in the binary Alpha Centauri. The~stars start at apoastron: A is in the top left corner, and~B is in the bottom right corner. From~left to right and top to bottom, we show the system at different evolutionary stages: 0, 0.5, 1, 1.5, 2.0, 2.5, 3, and~5 $P_{\rm b}$, where $P_{\rm b}$ is a binary orbital~period. The red dots represent the stars of Alpha Centauri. \label{fig:hydro-sky-evo}}
\end{figure}

Figure~\ref{fig:hydro-sky-evo} shows the evolution of the circumstellar discs in Alpha Centauri. The~system is shown as it would appear in the sky. We see that the discs experienced a strong gravitational perturbation when the stars were close to the periastron (see second and forth panels in the top row). Tidal spirals formed and material was exchanged as the stars got closer to each other. This happens because the outer disc radii are larger than the tidal truncation radii caused by the stars. After~a few orbits, this effect disappeared as the discs quickly reached their steady size set by tidal truncation. For~instance, after~three orbital periods, the~discs only experienced mild perturbations as the stars orbited around each other. For~the rest of the simulation, we did not see any noticeable morphological differences from one orbit to another---besides a very mild disc eccentricity excitation \citep{Martin2020}. \ms{The long-term stability of the discs is guaranteed by the gas, as it efficiently damps the eccentricity excitations within the discs (that would otherwise be triggered in pure N-body integrations). We expected the gaseous disc to start dissipating on time scales of the order of several Myr \citep{Manara+2023, Ribas+2015}. However, this effect did not affect the results presented in this work regarding disc dynamics.}

\begin{figure}[H]

\includegraphics[width=0.48\linewidth]{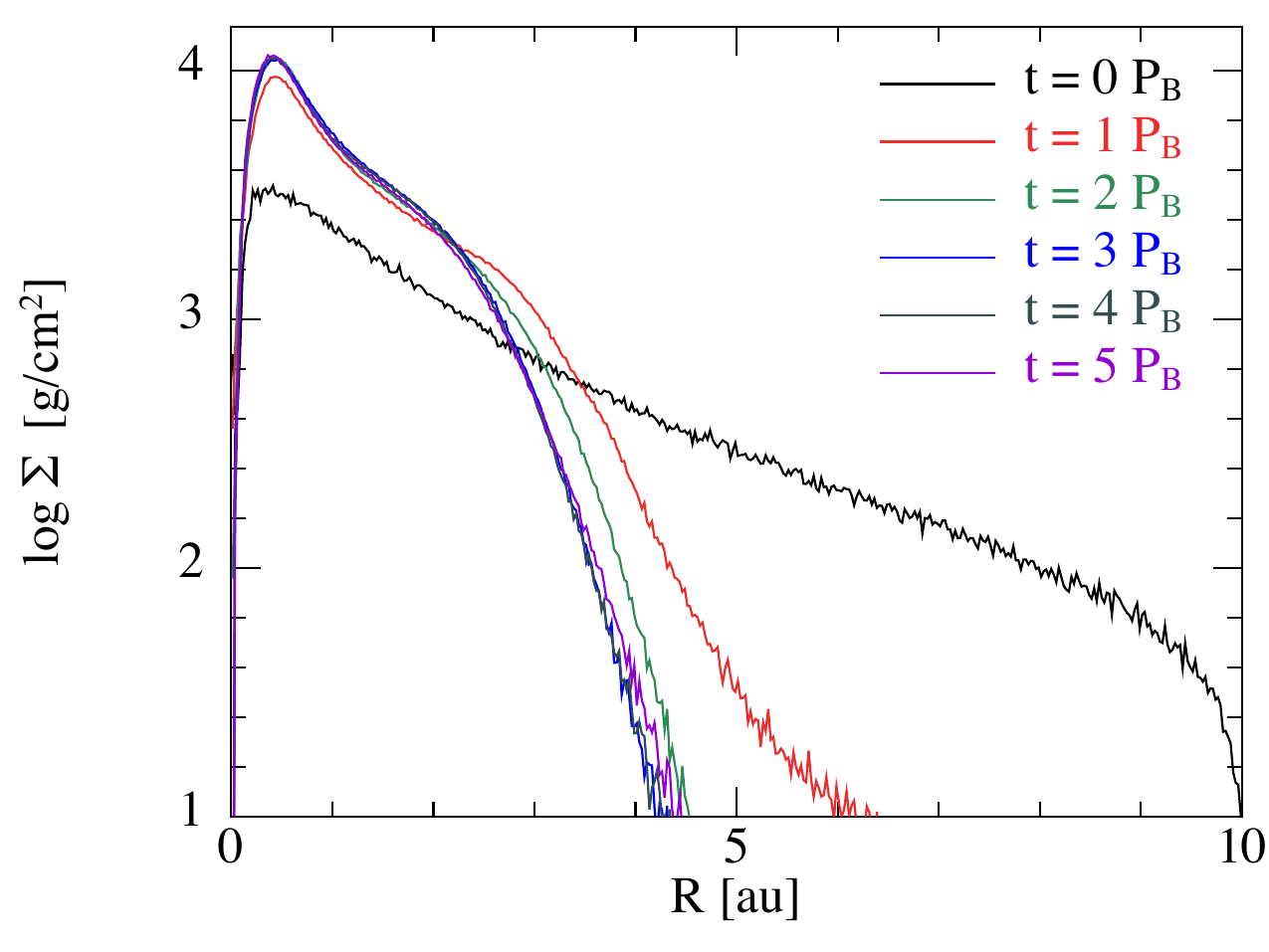}
\includegraphics[width=0.48\linewidth]{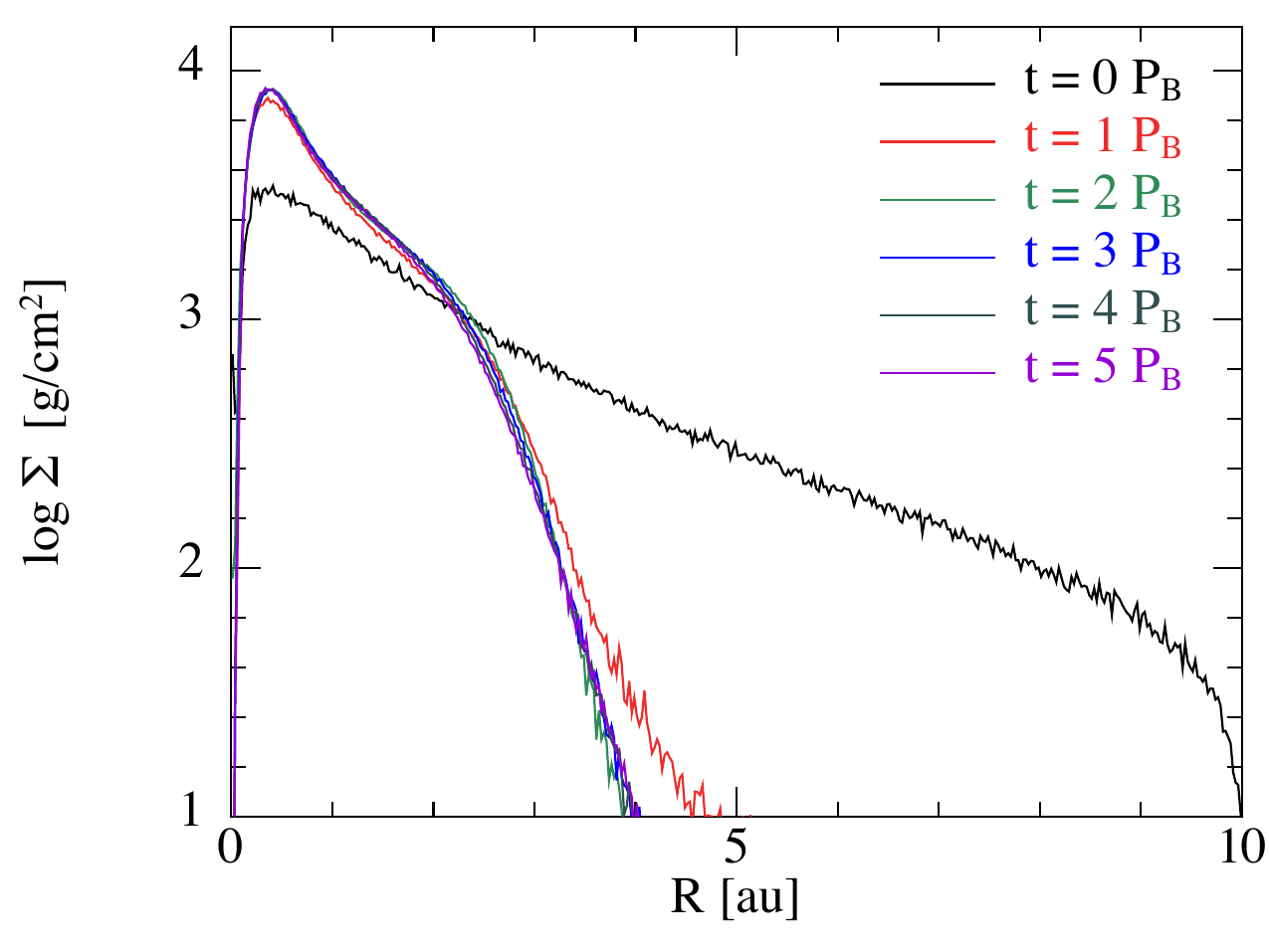}
\caption{Surface density profiles of the circumstellar discs around Alpha Cen A (\textbf{left}) and B (\textbf{right}). Different colours represent different evolutionary stages: 0, 1, 2, 3, 4, and~5 binary orbital~periods.\label{fig:surf-density}}
\end{figure}

Figure~\ref{fig:surf-density} exhibits the surface density profiles as a function of the distance to the central star for the circumstellar discs around A (left panel) and B (right panel). We see that the discs were initially extended up to 10~au following the density profile described in Section~\ref{sec:discs_setup}. After~only three binary orbital periods, both discs sizes remained practically unchanged. The~disc profiles around A and B were very similar and were marked by a drastic break at around 3~au from the host star. Material beyond that distance was less dense. In~other words, the~bulk of the disc was mainly contained below 3~au. We also note that, due to the difference in stellar mass, the~circumstellar disc around the secondary star B converged faster towards the tidal truncation radius compared to the disc around primary star~A.

The initial values used in this study to initialise the disc masses, although~in agreement with typical values \citep{Miotello2023}, were admittedly arbitrary and practically unconstrained from observations. We can, however, interpret our results in the broader context of circumstellar discs in binaries. Integrating the profiles in Figure~\ref{fig:surf-density} radially after five binary orbits, we find that the gaseous disc masses around A and B were equal to $3266$~$M_{\oplus}$ and $2167$~$M_{\oplus}$, respectively. Assuming a dust-to-gas ratio of 1\%, this means that there were about $33$~$M_{\oplus}$ and $22$~$M_{\oplus}$ of dust available to form rocky planets around A and B, respectively. The~rest of the material is either accreted by the stars or ejected from the system. In~other words, the disc around A is about 50\% more massive than the disc around B---although they had the same mass and size initially. As~expected, the~primary star A was able to retain more material around it with respect to the secondary star B \citep{Pichardo2005}. The~immediate consequence of this is that, if~we assume the same planet formation efficiency within both discs, then we should expect more massive planets around A compared to~B.

\begin{figure}[H]
\includegraphics[width=0.48\linewidth]{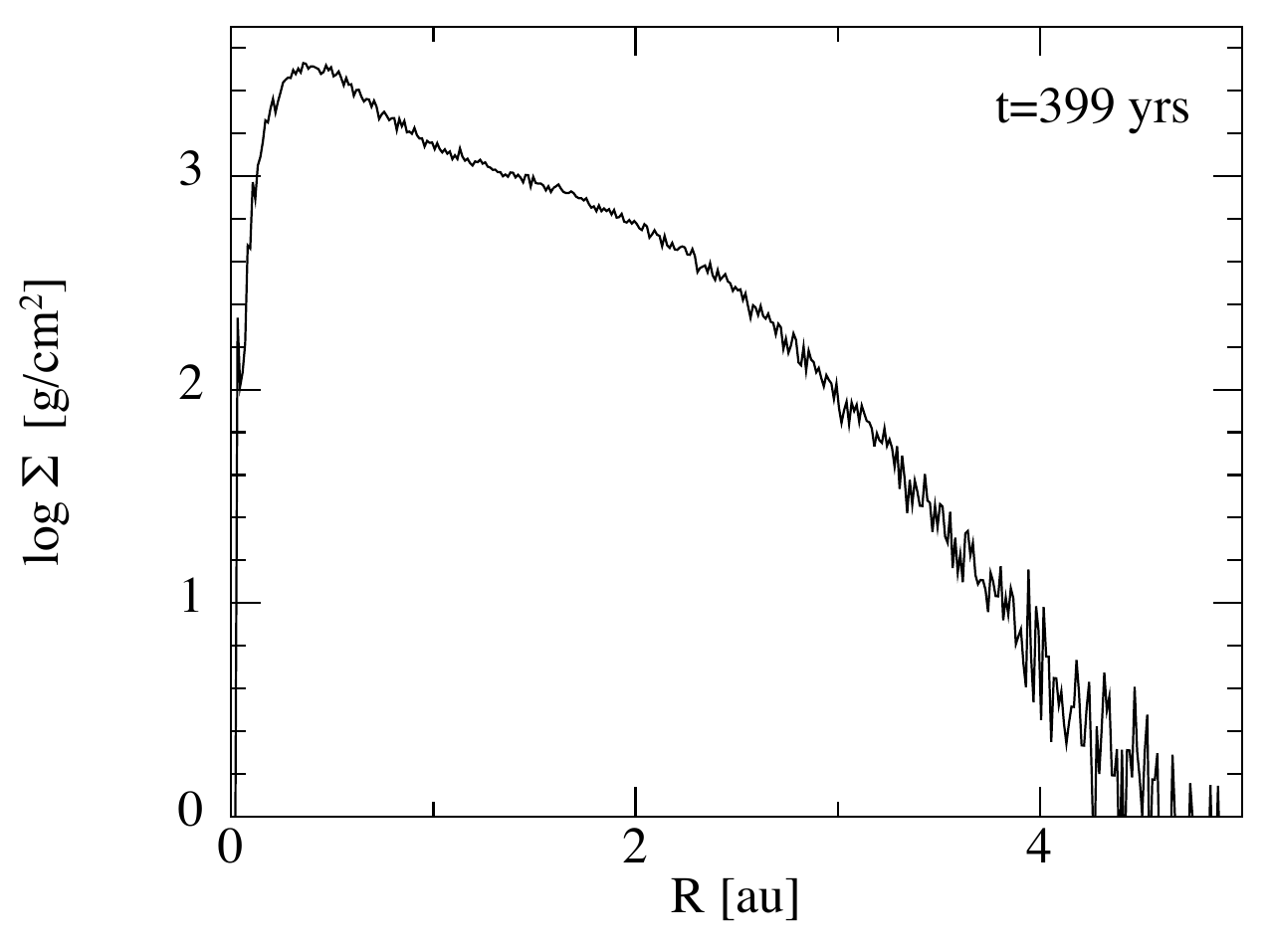}
\includegraphics[width=0.48\linewidth]{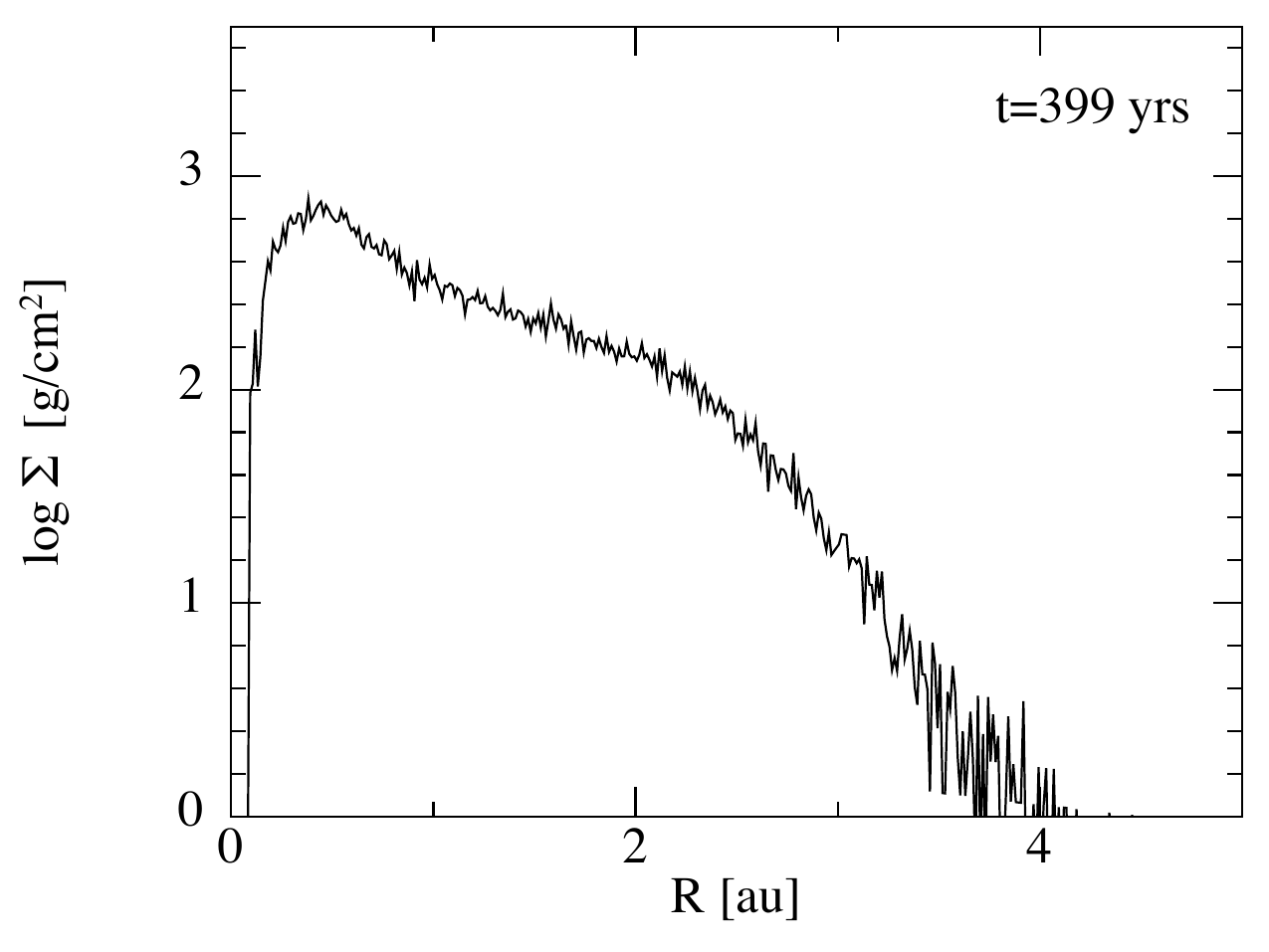}
\caption{Disc surface density profiles of the alien material captured around A from the disc around B (\textbf{left}) and~captured around B from the disc around A (\textbf{right})---after 5 binary orbital~periods.\label{fig:alien-surf}}
\end{figure}

\begin{figure}[H]
\includegraphics[width=0.495\columnwidth]{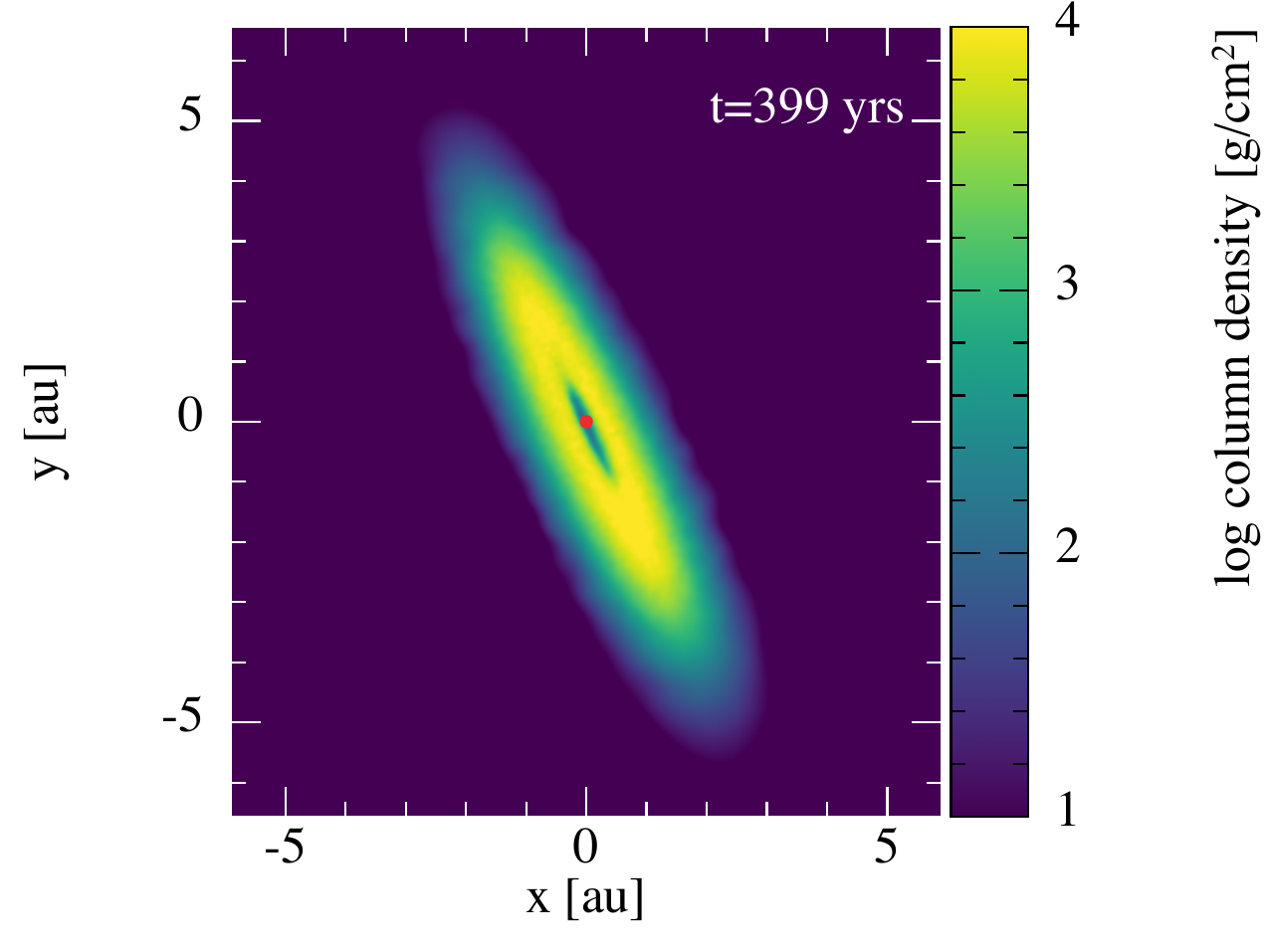}
\includegraphics[width=0.495\columnwidth]{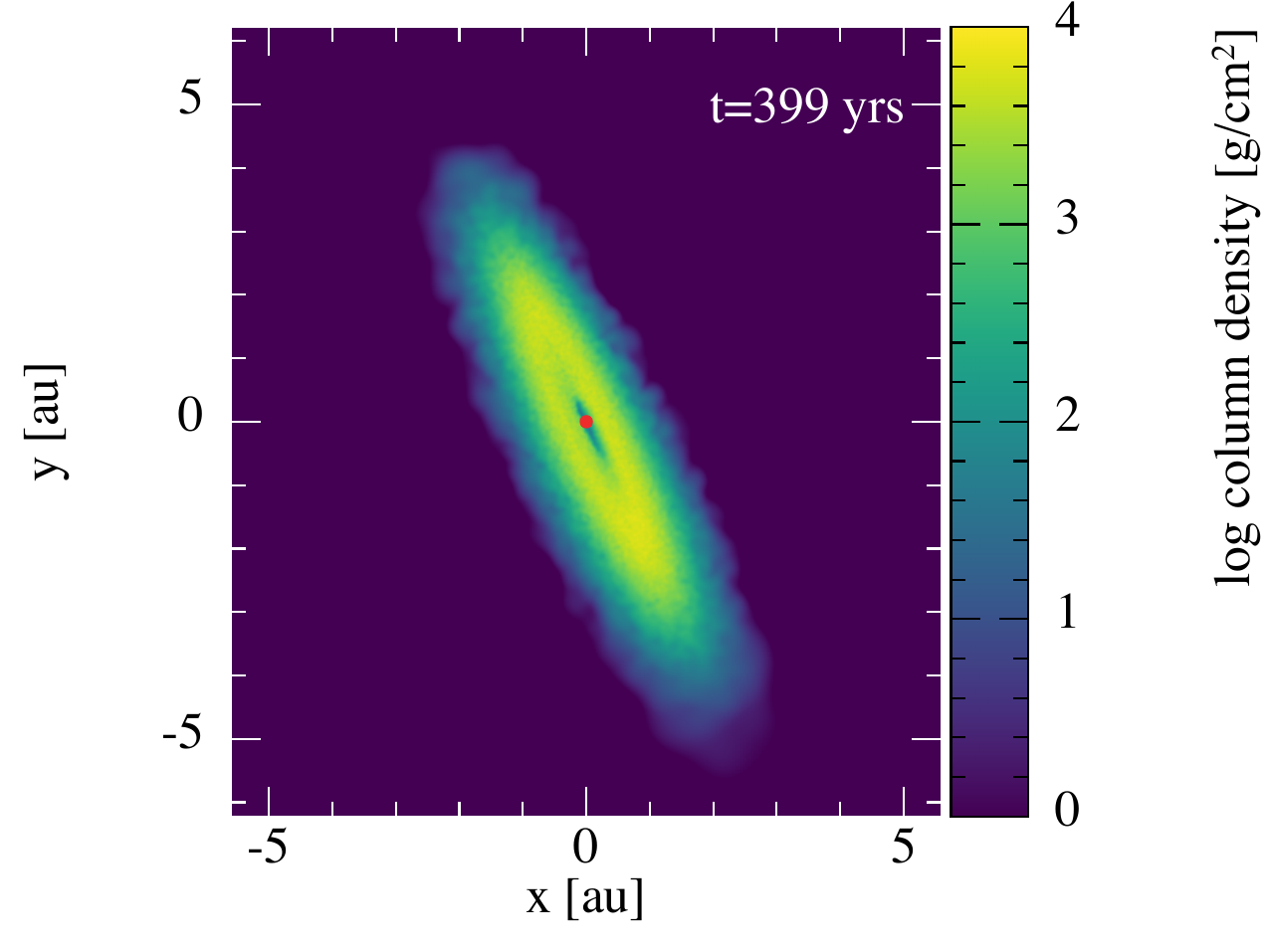}
\includegraphics[width=0.495\columnwidth]{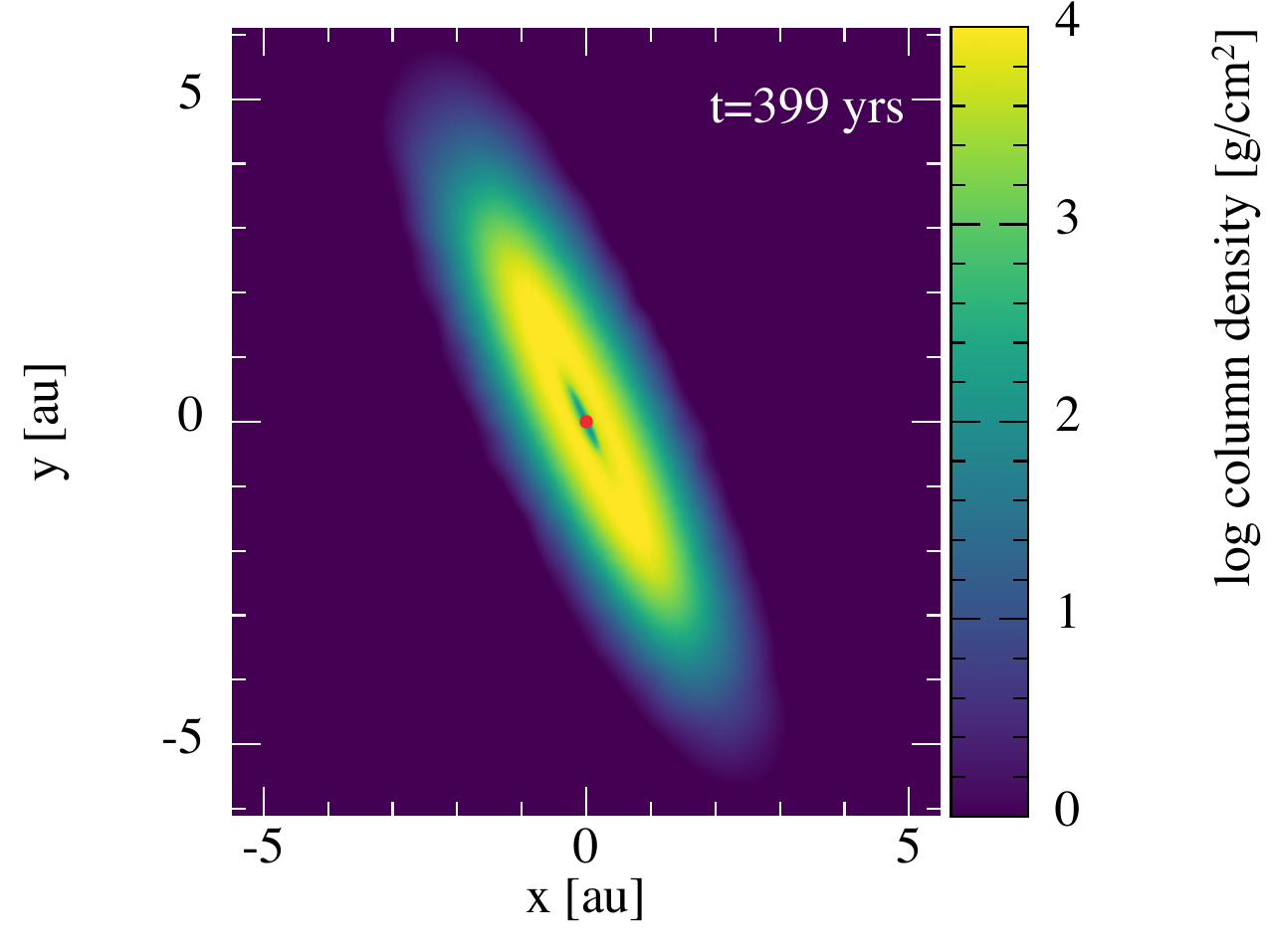}
\includegraphics[width=0.495\columnwidth]{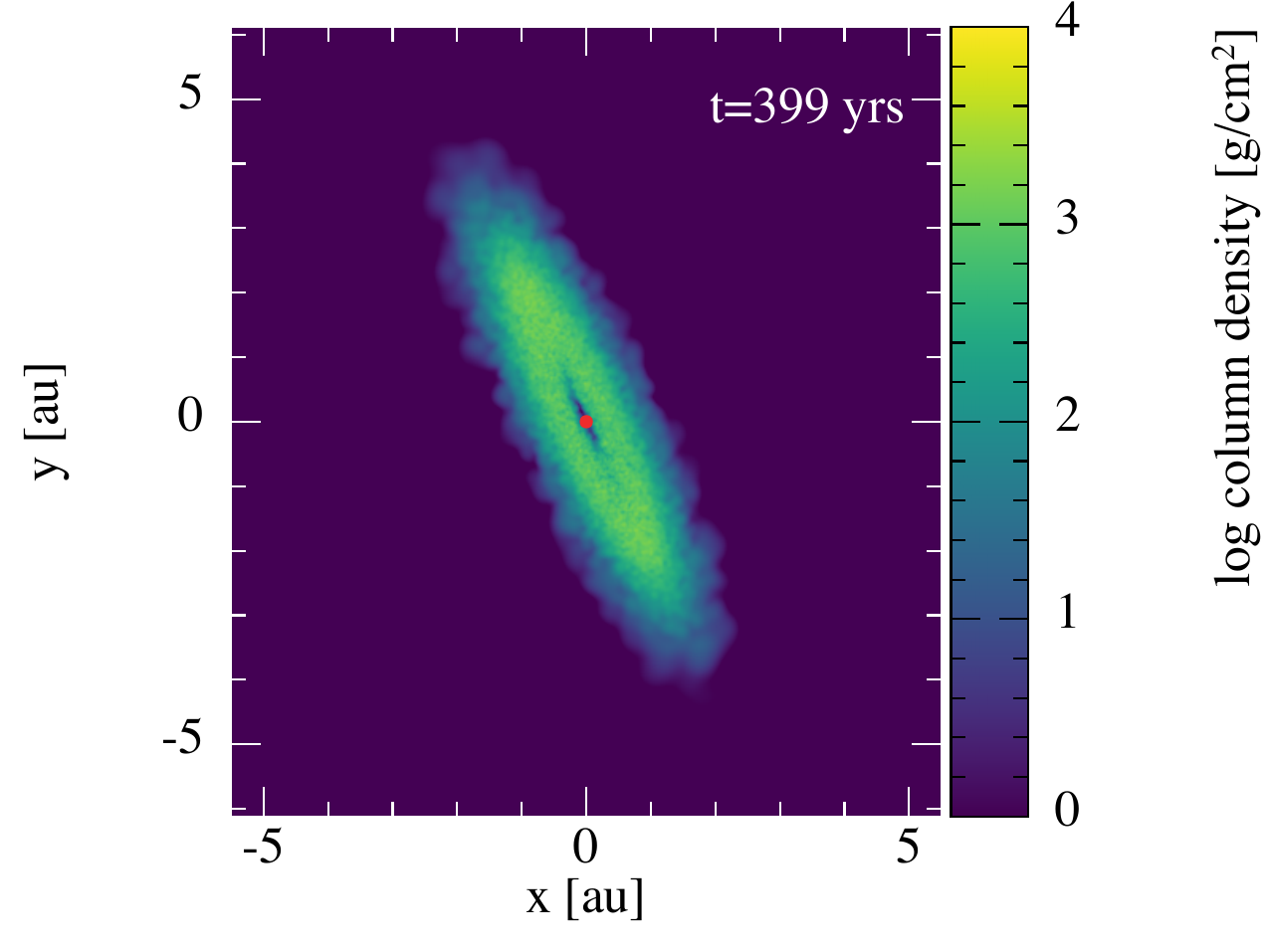}
\caption{Gas column density of the discs around A and B after 5 binary orbital periods. The~disc contents are split into pristine (\textbf{left column}) and alien captured material (\textbf{right column}). \textbf{Top left} (\textbf{bottom left}) panel shows the material that was around A (B) initially. Top right (\textbf{bottom right}) panel shows the alien captured material by A (B) from B (A). The~captured (alien) material is shown to be well distributed and mixed with the disc pristine~material. The red dots represent the central stars. \label{fig:alien-gas}}
\end{figure}

It is also interesting to note that the circumstellar discs exchanged gaseous material with each other during the first orbital periods. This is mainly due to the initial disc sizes we considered. However, when the binary Alpha Cen formed, it is not unreasonable to consider that the circumstellar discs were more radially extended than the tidal disc truncation radii of about 3~au. At~any rate, we measured the amount of (alien) material that was transferred from disc A to disc B and vice~versa. Figure~\ref{fig:alien-surf} shows the surface density profiles of the material that was initially orbiting around one star and was transferred to the other star's disc. First, we note that there was more alien material around the primary A than around the secondary B, given that the disc around B loses gas more easily. Overall, the~surface density values of the alien material were an order of magnitude below those of the entire disc. This indicates that these discs were mildly polluted by gas coming from the other disc. The~left column of Figure~\ref{fig:alien-gas} shows the gas distributions of the material around A (top panel) and B (bottom  panel) that were initially around A and around B, respectively---i.e., the~captured material is not plotted. Instead, the~right column shows the gas distributions of the alien material captured by A from B (top) and~of the alien material captured by B from A (bottom). All these figures show the discs after five binary orbital periods, since we did not see meaningful variability for later evolutionary stages. The~implications of this initial transfer of material between circumstellar discs are discussed in Section~\ref{sec:discussion}. \ms{Last, we note that the formation of individual circumstellar discs from a circumbinary gas reservoir is beyond the scope of this work. This process has been recently studied in the context of misaligned young binaries \citep{Price+2018, Smallwood+2023}. Alpha Cen could have hosted a circumbinary disc in the past, but~we currently lack evidence pointing in that direction.}

\unskip

\unskip

\section{Stability of Planets in Alpha~Centauri}
\label{sec:stability}

We now supposed that planets have formed in Alpha Cen, and we employed numerical tools to evaluate the stability of these bodies within the triple stellar system. Such planets are subject to intricate perturbative interactions that can challenge their long-term stability. Although~the process of planetary formation in this kind of system remains not fully understood, our preceding sections have demonstrated that substantial portions of circumstellar disc material could be available for planetary formation, thus potentially favouring the emergence of low-mass planets. In~this section, we shall delve into the dynamic stability of these systems by employing the Mean Exponential Growth of Nearby Orbits ({\tt MEGNO}) stability parameter (see, e.g.,~\cite{Cincotta2000}).

The {\tt MEGNO} parameter is a powerful numerical tool for assessing the stability of N-body gravitational systems. It quantifies the exponential growth of nearby orbits in phase space over a specified time span. In~essence, {\tt MEGNO} gauges the rate at which adjacent trajectories in phase space diverge over time, thereby offering a quantitative metric for orbital stability. The~{\tt MEGNO} algorithm integrates the equations of motion for a set of closely aligned initial conditions and computes their rate of divergence. A~{\tt MEGNO} value approaching two corresponds to a regular orbit and, hence, a stable system, whereas values substantially larger than two imply chaotic orbits and potentially unstable systems. In~this work, we used the {\tt MEGNO} implementation available in the {\tt REBOUND} package \cite{rebound1,rebound2,rebound4}.

Our experimental approach was as follows: first, we placed massless test particles uniformly distributed in semimajor axis and eccentricity around each of the stars, and we evaluated the {\tt MEGNO} parameter. This allowed us to identify the system's allowed and forbidden resonant regions. Then, we placed particles with one Earth mass to ascertain which regions remained viable for hosting Earth-like planets. Last, we introduced individual planets with masses which followed the observed rocky planet mass distribution in our Solar System. By~doing so, we created planet analogues of the inner Solar System but within a triple star environment. It is worth mentioning that planets in triple star systems can have stable inclined orbits, e.g.,~\citep{Busetti2018}. In~this work, however, we limited ourselves to the coplanar case (i.e., the~planet is coplanar with the inner binary system). Incidentally, this configuration corresponds to more stable circumstellar orbits in double and triple~systems.

\subsection{Stability of Test Particles in Alpha~Centauri}

We note $a$ and $e$ as the particle semimajor axis and eccentricity, respectively. The~inclusion of test particles on~S-type orbits in the $a$--$e$ parameter space constitutes a preliminary estimate of the stability within the binary system. We therefore placed massless particles around each of the stars in the Alpha Cen binary. The~orbital initialisation in {\tt REBOUND} was the same as the one described in Section~\ref{sec:bin_setup}. For~this simulation, we restricted the semimajor axis of the test particles to range from 0.1 to 5~au, which was based on the estimated extension of the circumstellar discs after their interaction (see Figure~\ref{fig:surf-density}). Furthermore, we did not consider eccentricities over 0.5, since this would hamper planet stability and also prevent additional planets in the system \citep{Quarles2016}. We set the maximum integration time to  10,000 years, which corresponds to about 127 orbits of the inner binary system ($P_{\rm b} \approx 80$~yr). For~these simulations, we used the symplectic integrator 'whfast', with~a time step of $1 \times 10 ^{-5}$ yr, which allowed us to map the orbit closest to the star with at least 100~points.

\begin{figure}[H]
\includegraphics[width=0.5\columnwidth]{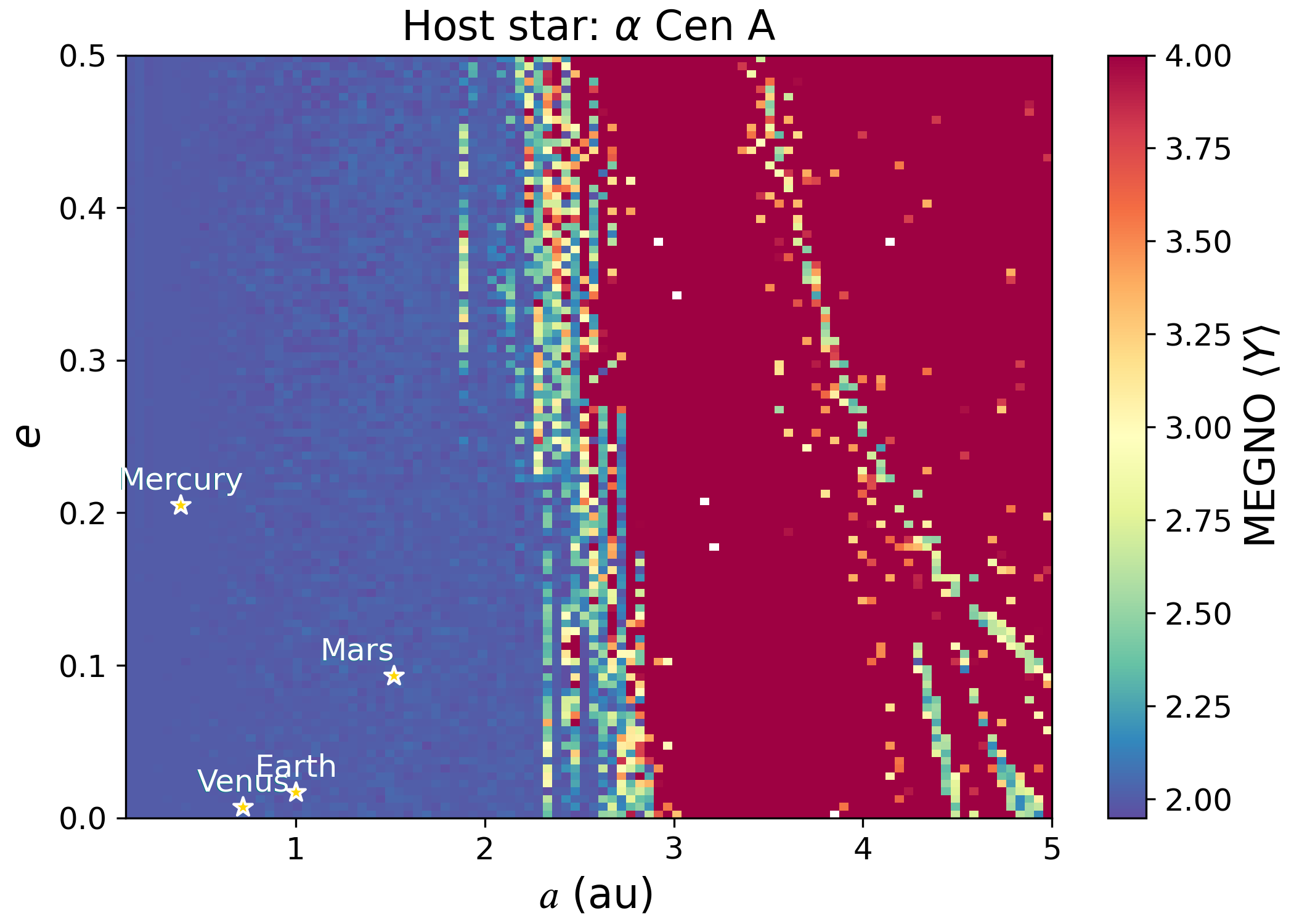}
\includegraphics[width=0.5\columnwidth]{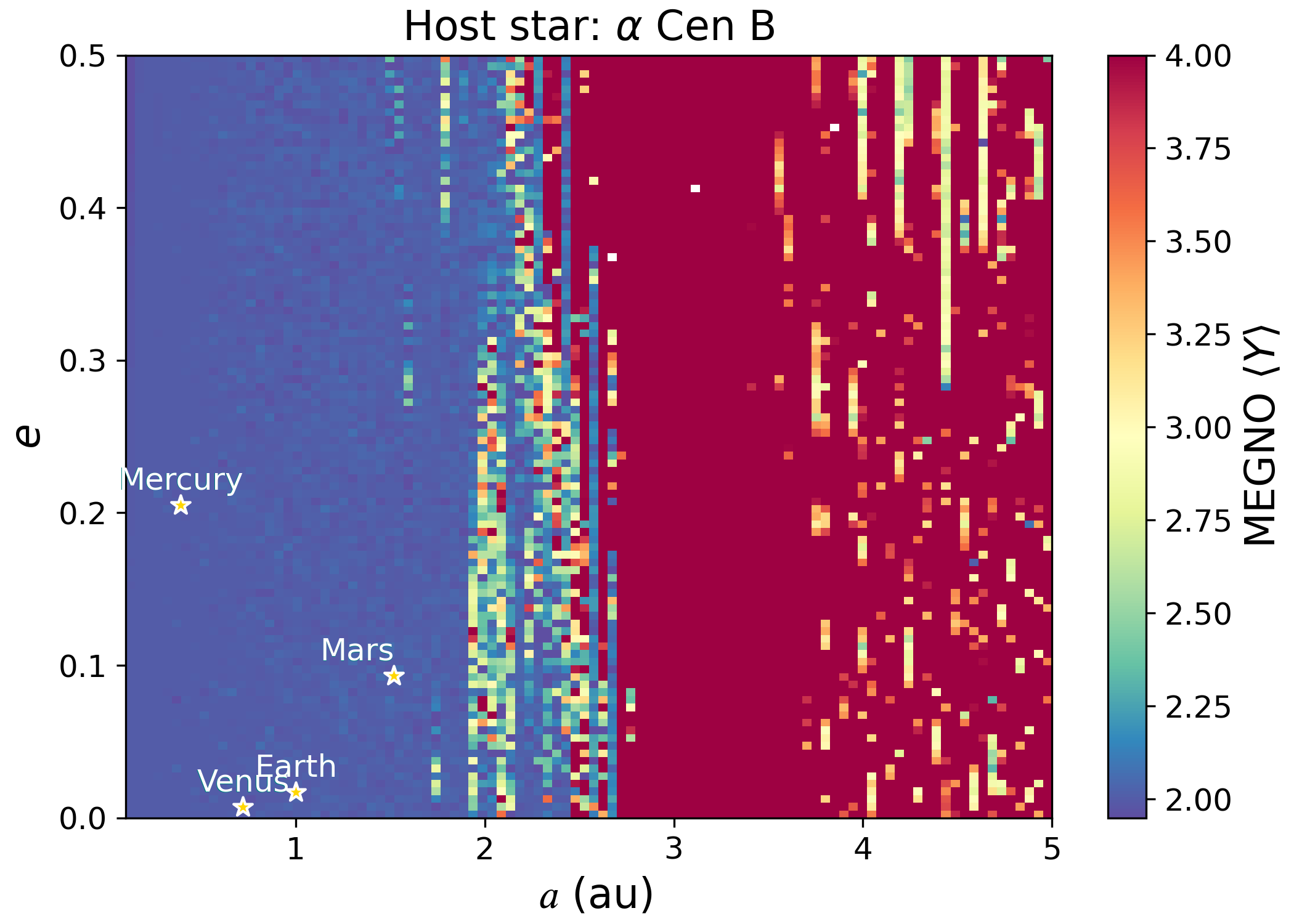}
\caption{Stability maps based on {\tt MEGNO} values for test particles over a 10,000-year period. Dynamically stable regions are coloured in~blue. \label{fig:megnom0}}
\end{figure}

Figure~\ref{fig:megnom0} presents the results of such simulations. The~regions coloured in blue correspond to zones in the $a$--$e$ parameter space, whose stability is assured by the algorithm, while the red zones are regions of high instability. For~scale purposes, we have added the rocky planets of the Solar System to each graph. The~blue areas in the left (right) panel correspond to the stability regions for test particles around star A (B). We identified two distinct regions in terms of stability around A: one spanning from 0.1 to $\sim$2.5 ~au and~the other from $\sim$2.5 to 5~au. The~former is narrower and contains all the stable orbits for $e$ ranging from $0$ to $0.5$. Remarkably, individual massless particles with the same orbital parameters as those of the inner planets in the Solar System would maintain regular orbits. A~similar scenario was observed around star B, where the parameter space was segmented again into two zones according to stability: one from 0.1 to 2.5~au and~another beyond 2.5~au. Here, the~latter region was highly unstable and inhospitable for test particles. In~the case of star C (shown in Appendix~\ref{sec:appendix_stab}), owing to its considerable distance from the binary system, the~entire range of the parameter space remained perfectly stable for the particles under consideration. Hence, individual massless particles with the same orbital parameters as the planets of the inner Solar System should be stable around Alpha Cen A, B, and~C.

\unskip

\subsection{Stability of Earth-like Planets in Alpha~Centauri}

\begin{figure}[H]
\includegraphics[width=0.5\columnwidth]{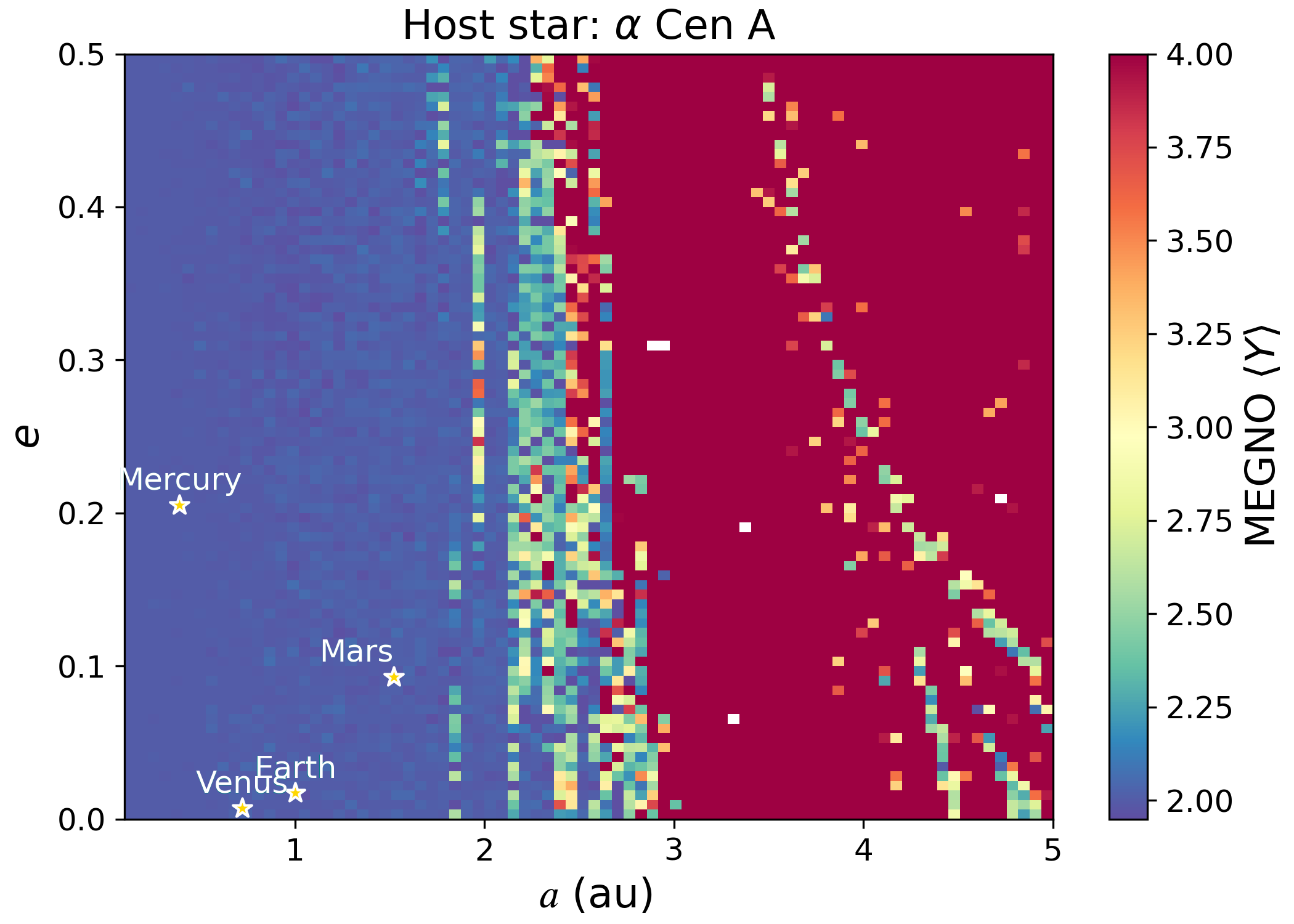}
\includegraphics[width=0.5\columnwidth]{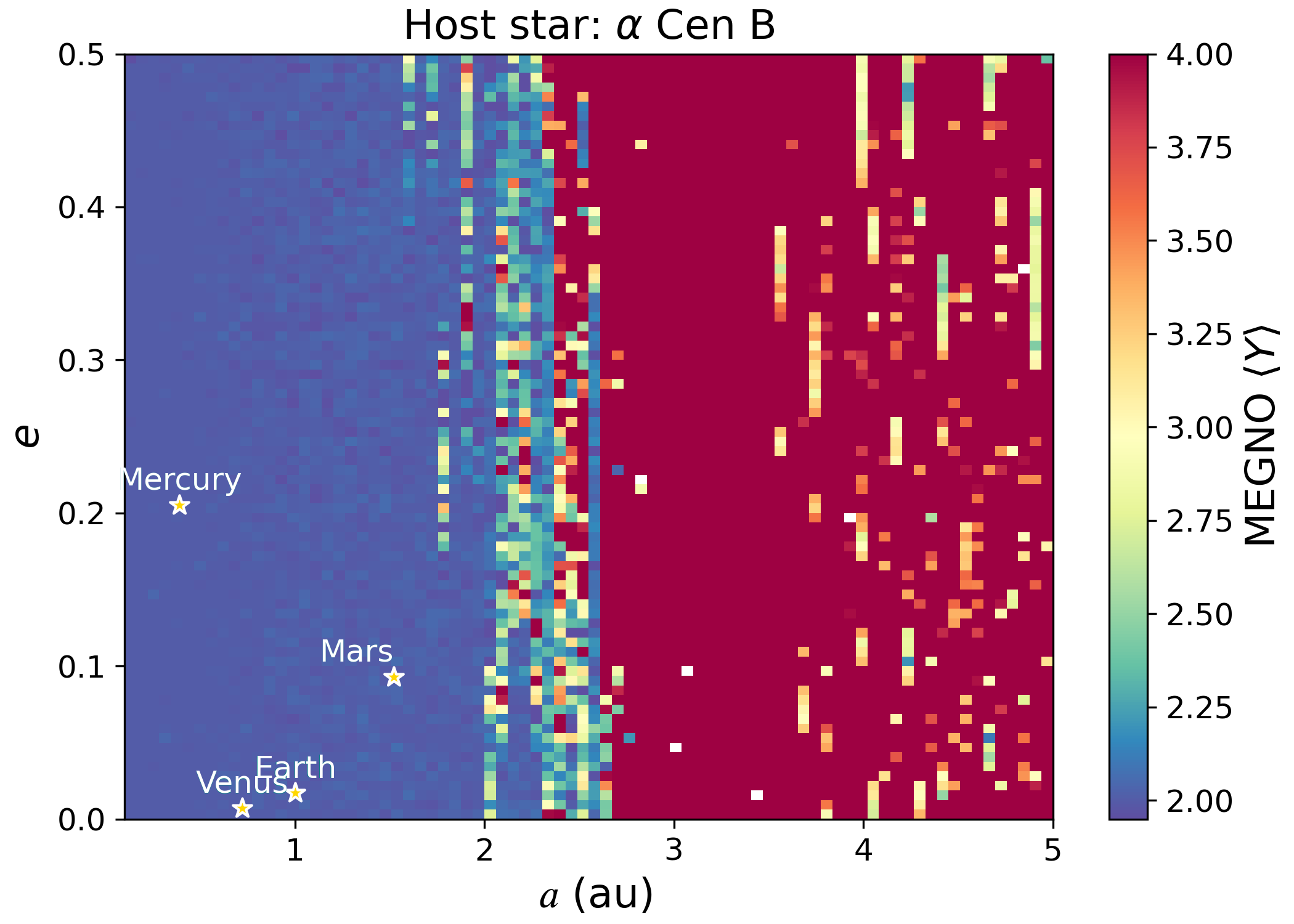}
\caption{Stability maps based on {\tt MEGNO} values for 1 Earth-mass planet over a 10,000-year period. Dynamically stable regions are coloured in~blue. \label{fig:megnom1}}
\end{figure}

The fact that test particles cannot thrive in certain regions of a triple star system, which are too hostile and unstable, does not necessarily imply that massive planets cannot survive. Indeed, the~very mass of the planet in question can slightly modify the inner binary parameters or~even alter the external companion orbit through gravitational perturbations. This could potentially lead to more stable conditions, which would modify the resulting maps shown in Figure~\ref{fig:megnom0}. In~this subsection, we consider massive (rocky) planets that can interact with the three stars of Alpha Cen. Using the same orbital parameters as in the previous section, we obtain the maps shown in Figure~\ref{fig:megnom1} for Earth-like~planets.

For 1~$M_\oplus$ planets, the~inner stability region around A remained practically unchanged---with a stability boundary at around 3~au for circular planets and around 2.5~au for eccentric planets ($0<e\leq0.5$). We observed the same phenomenon for 1~$M_\oplus$ planets around B, with a narrow band up to around 2~au, which contains all the massive planets with stable orbits. All the planets beyond that limit were unstable. Given that our revised {\tt MEGNO} maps for 1~$M_\oplus$ planets remained the same compared to the massless case, we conclude that the planet mass has a limited effect on stability for rocky~planets.

\unskip

\subsection{Stability for Inner Solar System Analogues in Alpha~Centauri}
\label{sec:stab_ss_mass}

\begin{figure}[H]
\includegraphics[width=0.75\columnwidth]
{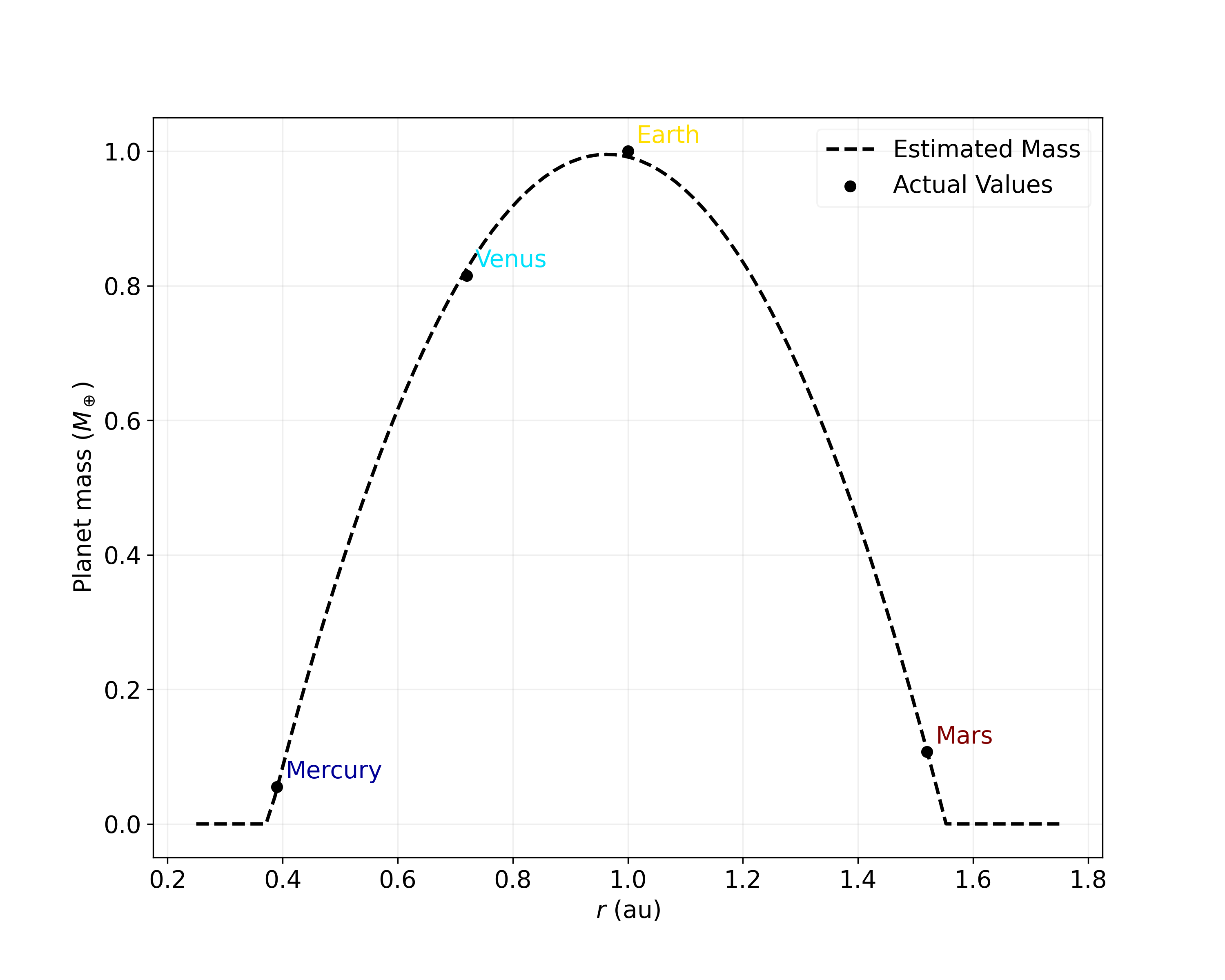}
\caption{Estimated vs. real mass of rocky planets as a function of~distance.\label{fig:planmass}}
\end{figure}

In this section, we study the stability of Solar System planetary analogues orbiting around stars A, B, and~C in Alpha Centauri. To~set the planetary masses as a function of the distance from the host star, we use an empirical relation based on a second-order polynomial equation. This relation is calibrated based on the inner Solar System's masses and has the following expression:
\begin{linenomath}
\begin{equation}
    m(r) = -2.8664 \cdot r^2 + 5.5251 \cdot r - 1.6671,
\end{equation}
\end{linenomath}
where \( m \) is the planet mass in Earth masses, and $r \in [0.35, 1.9]$ represents the distance from the host star in au (see Figure~\ref{fig:planmass}). This oversimplified function offers a useful method for estimating the mass of hypothetical Solar System-like planets in the Alpha Centauri system. The~polynomial function is a mere fitting of four data points, thereby corresponding to the inner rocky planets of the Solar System: Mercury, Venus, Earth, and~Mars. Consequently, this model does not possess a robust physical justification and should be utilised with~caution.

\begin{figure}[H]
\includegraphics[width=0.5\columnwidth]{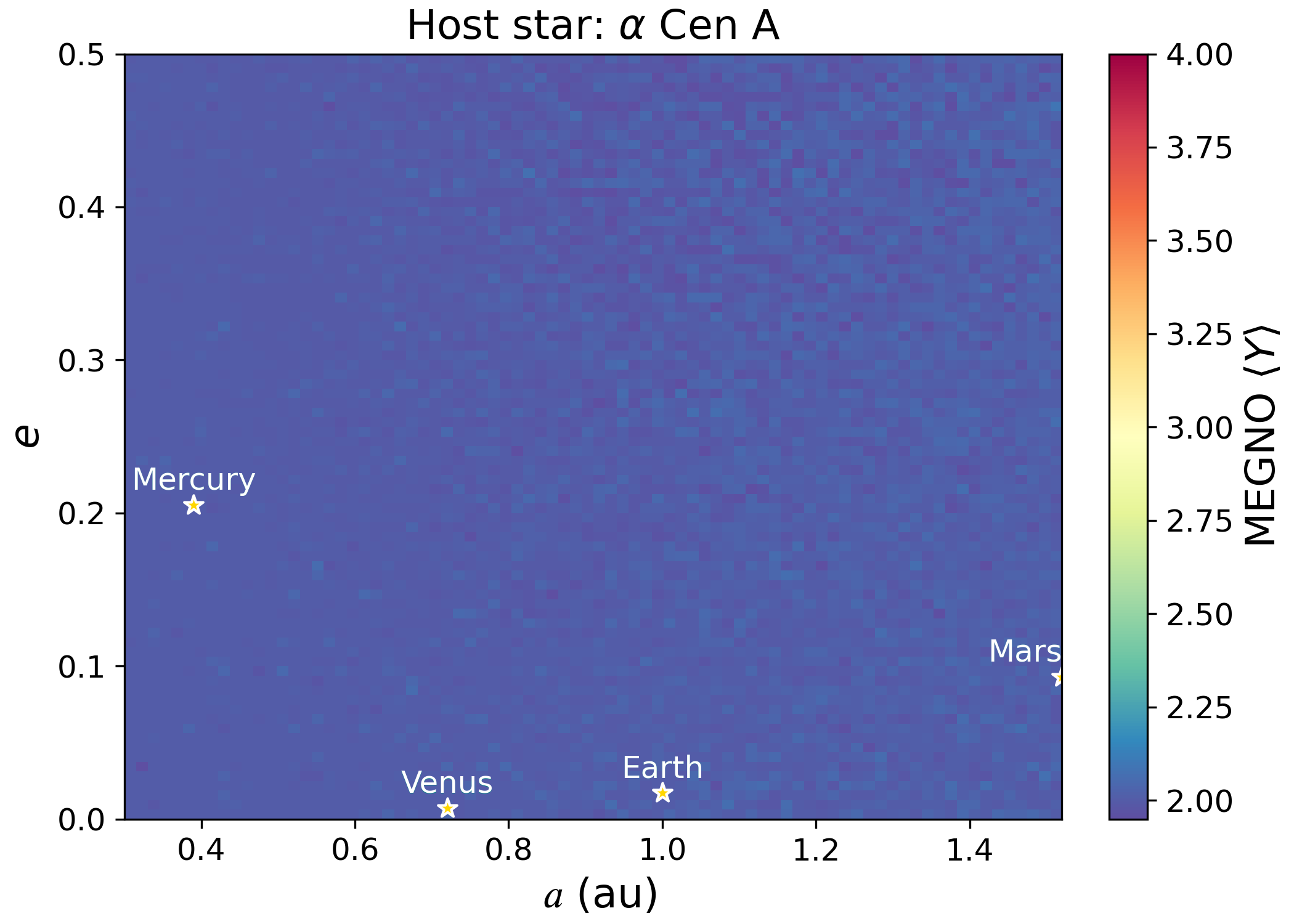}
\includegraphics[width=0.5\columnwidth]{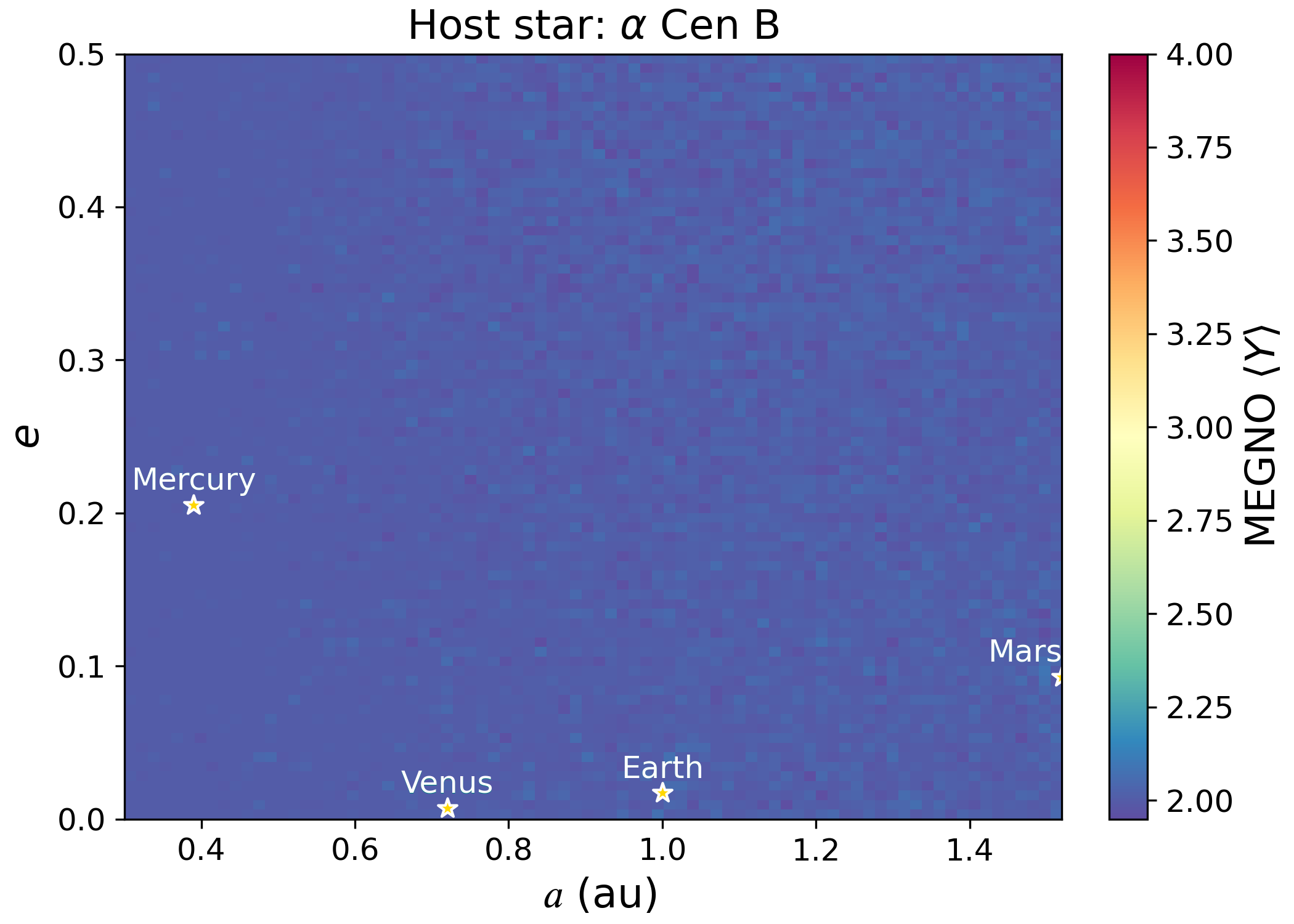}
\caption{Stability maps based on {\tt MEGNO} values for the inner Solar System analogues over a 10,000-year period. Dynamically stable regions are coloured in~blue. \label{fig:megno_ss}}
\end{figure}

Figure~\ref{fig:megno_ss} illustrates the {\tt MEGNO} map on the $a$--$e$ plane for this kind of planets for~semimajor axes ranging between $0.3$ and $1.5$~au. For~the considered timescale ($10,000$~yr), these celestial bodies exhibited perfect regularity in the region under study. Therefore, stars A and B could host inner Solar System analogues on stable orbits. It is worth reminding that the stability maps presented here were done for four bodies only: three stars and a planet. The~initial conditions for the planet's semimajor axis and eccentricity varies within specified values, arranged in a 100 $\times$ 100 grid. In~essence, to~construct each map, 10,000 integrations of the system were required, thus spanning a duration of 10,000 years. The~stability analysis of the multiplanetary systems in Alpha Cen using the {\tt MEGNO} indicator is left for future work. The~implications of these results are discussed further in Section~\ref{sec:discussion}.

\begin{figure}[H]
\includegraphics[width=0.5\columnwidth]{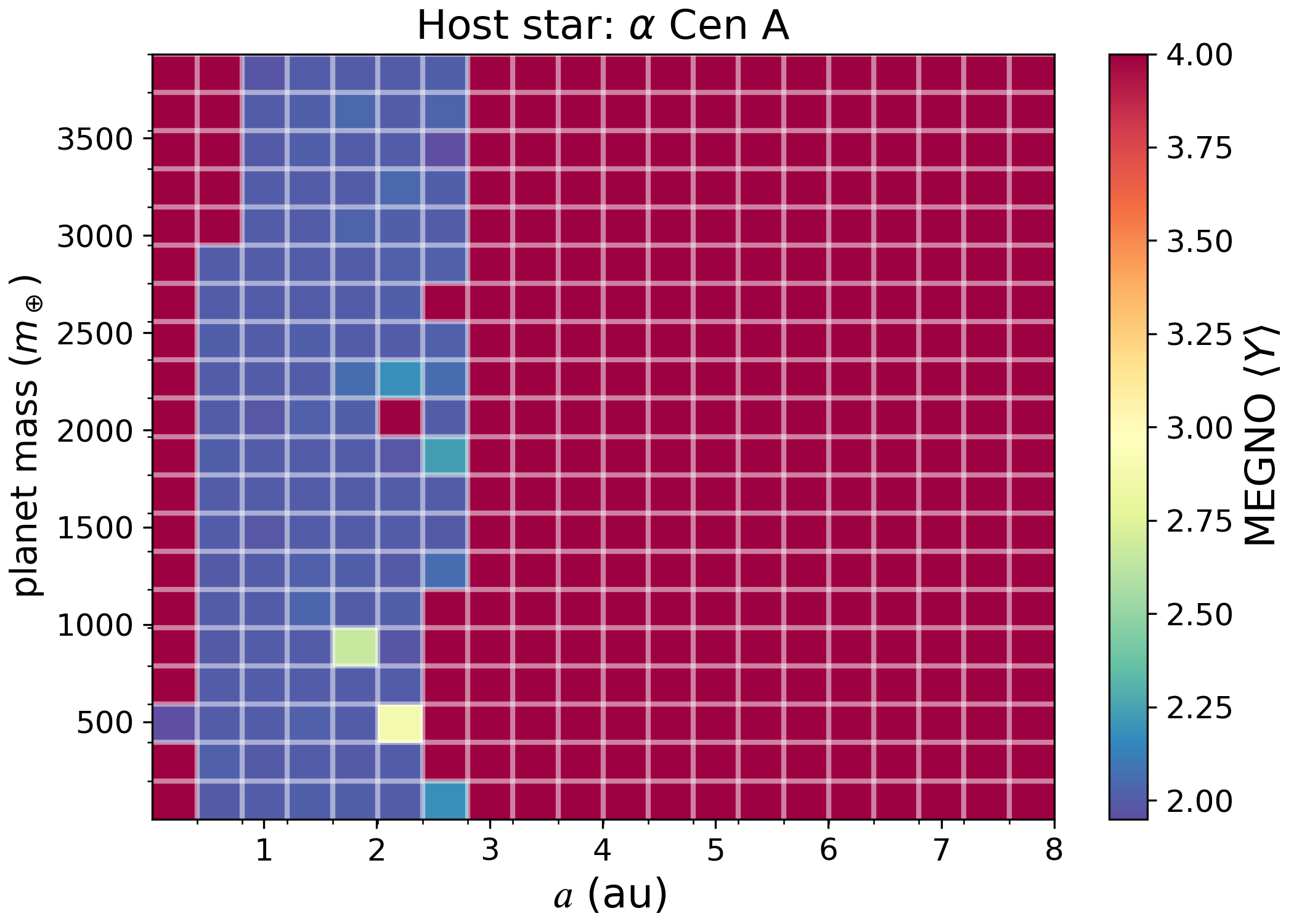}
\includegraphics[width=0.5\columnwidth]{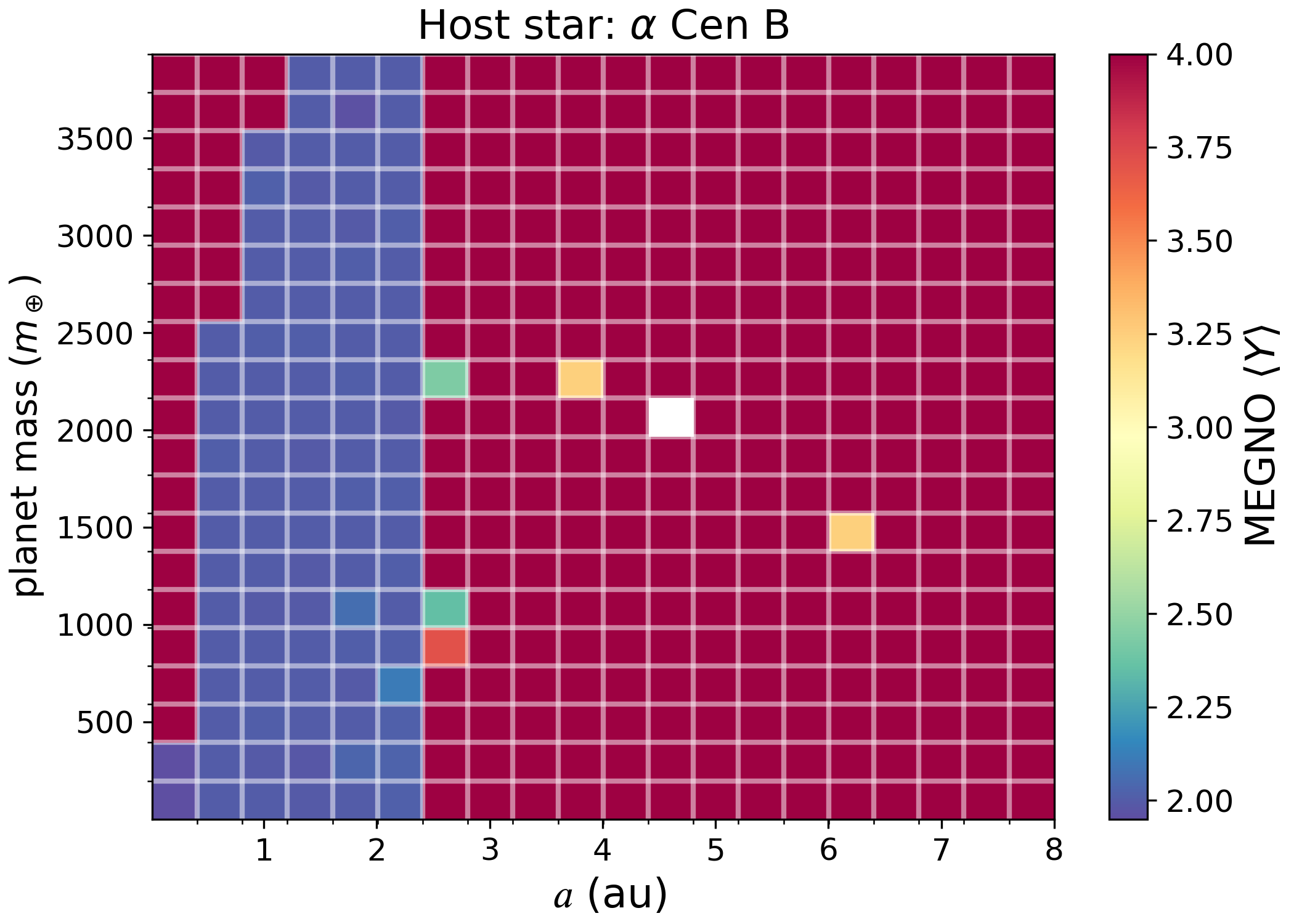}
\caption{\ms{Maps of dynamical stability assessed using {\tt MEGNO} values for the semimajor axis (\(a_p\)) and planetary mass (\(m_p\)) space over a period of 100,000 years. Regions indicating dynamical stability are highlighted in~yellow.} \label{fig:megnom_a_vs_m}}
\end{figure}

\ms{Finally, we examined the stability of a system with a single circumstellar planet around each star, thus focusing on the planetary mass and its semimajor axis. The~map in Figure~\ref{fig:megnom_a_vs_m} displays {\tt MEGNO} results for 10,000-year-long simulations of planets on circular orbits (using a 20 $\times$ 20 grid). Dark and copper areas correspond to unstable and stable regions, respectively. Our findings suggest that there is no maximum mass limit in the region spanning approximately from 0.5 to 2.5 au, thereby indicating that planets of varied masses should be stable in this area of the parameter space.}

\unskip   
\section{Detectability of Planets in Alpha~Centauri}
\label{sec:detectability}

The detection of extrasolar planets has been tremendously successful over the last thirty years, with~more than 5000 confirmed exoplanets\endnote{\url{https://exoplanet.eu/catalog/} (accessed on 17 January 2024)} to date. This achievement was possible thanks to the careful study of light and radial velocity (RV) curves with reported anomalous dips or modulations. Other techniques such as direct imaging and gravitational lensing methods have also contributed to increasing the exoplanet demographics, but~in lower numbers. Owing to its proximity to Earth, the~Alpha Centauri system has been subjected to an intensive and scrupulous search. So far, only one planet has been confirmed around Proxima Centauri \citep{Anglada2016}---although several candidates have been proposed~\cite{Damasso2020, Wagner2021, Faria2022}.

\subsection{Radial~Velocity}
\label{sec:radvel}
If we assume that hypothetical planets around A and B are coplanar with the binary orbital plane, then the inclination of the binary orbital plane (of about $79\deg$ with respect to the sky) significantly lowers the probability of planetary transits. Here, we are mainly interested in exoplanets in the binary Alpha Cen AB. Therefore, in~this section, we assess the detectability of inner Solar System analogues around A and B (see Section~\ref{sec:stab_ss_mass}) through radial velocity methods. Our goal is to compare our synthetic RV curves against the capabilities of current detectors such as HARPS\endnote{\url{https://www.eso.org/public/teles-instr/lasilla/36/harps/} (accessed on 17 January 2024)}, NEID\endnote{\url{https://neid.psu.edu/} (accessed on 17 January 2024)}, and~ESPRESSO\endnote{\url{https://www.eso.org/sci/facilities/paranal/instruments/espresso.html} (accessed on 17 January 2024)}.

\begin{figure}[H]
\begin{adjustwidth}{-\extralength}{0cm}
\centering 

\includegraphics[width=0.43\columnwidth]{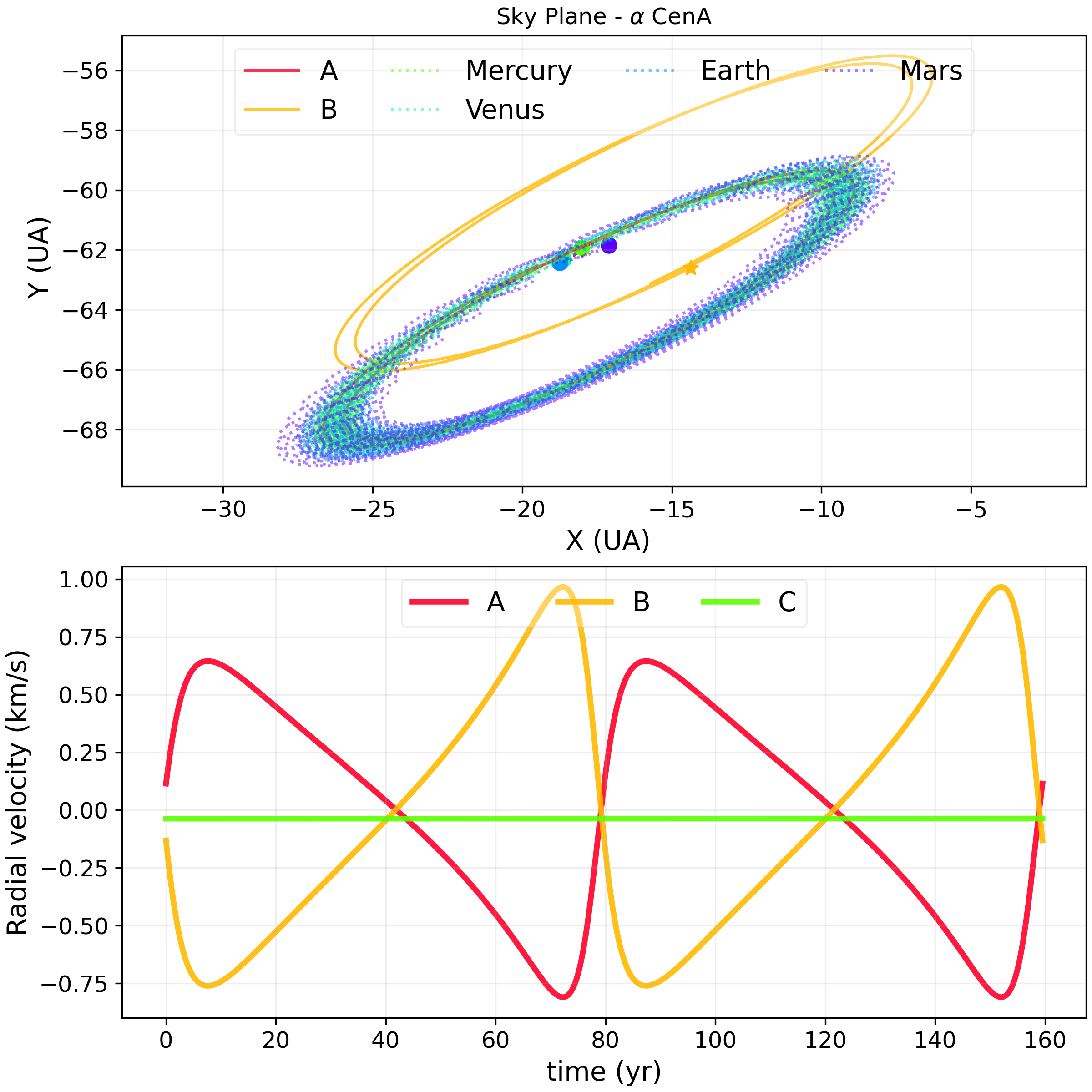}
\includegraphics[width=0.43\columnwidth]{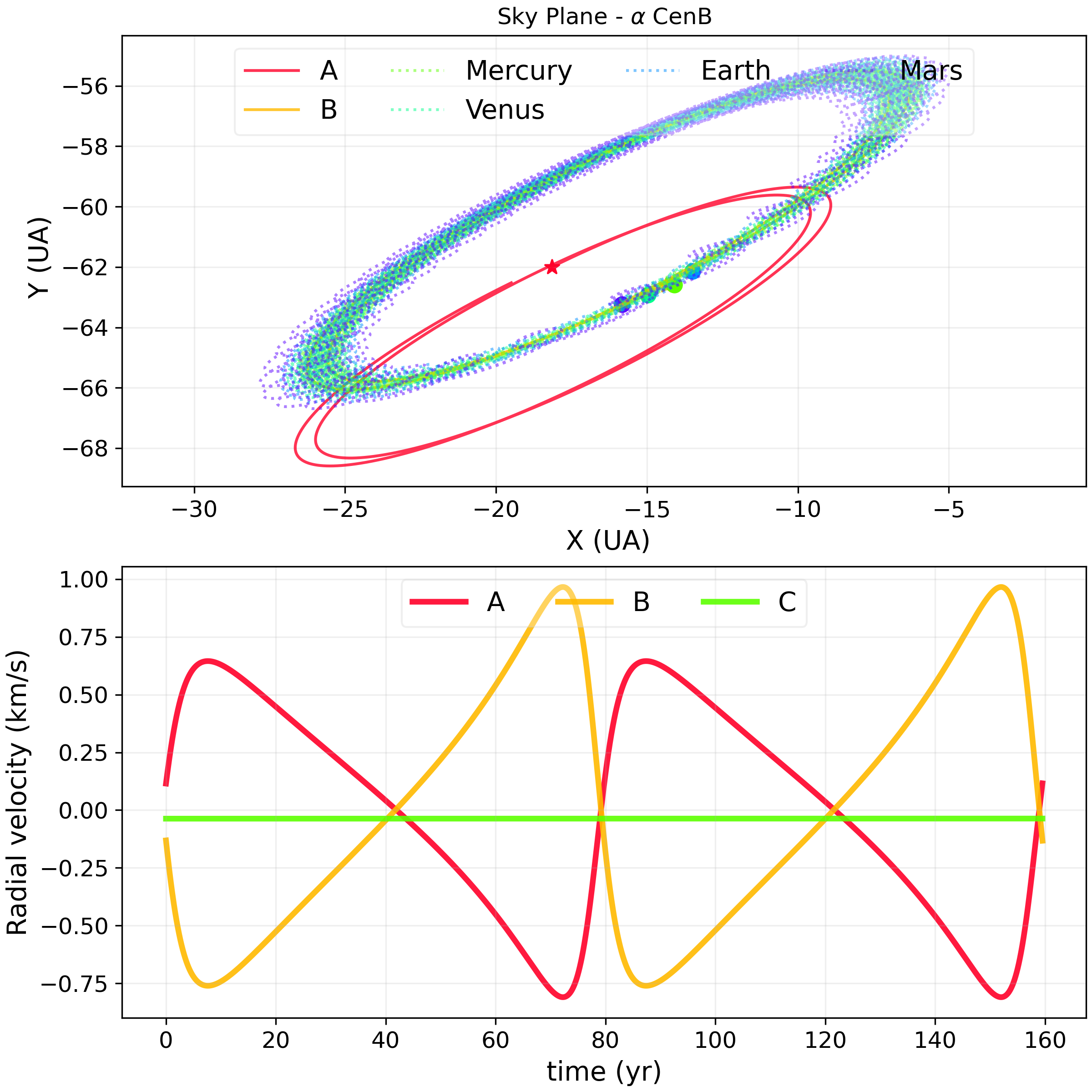}
\includegraphics[width=0.43\columnwidth]{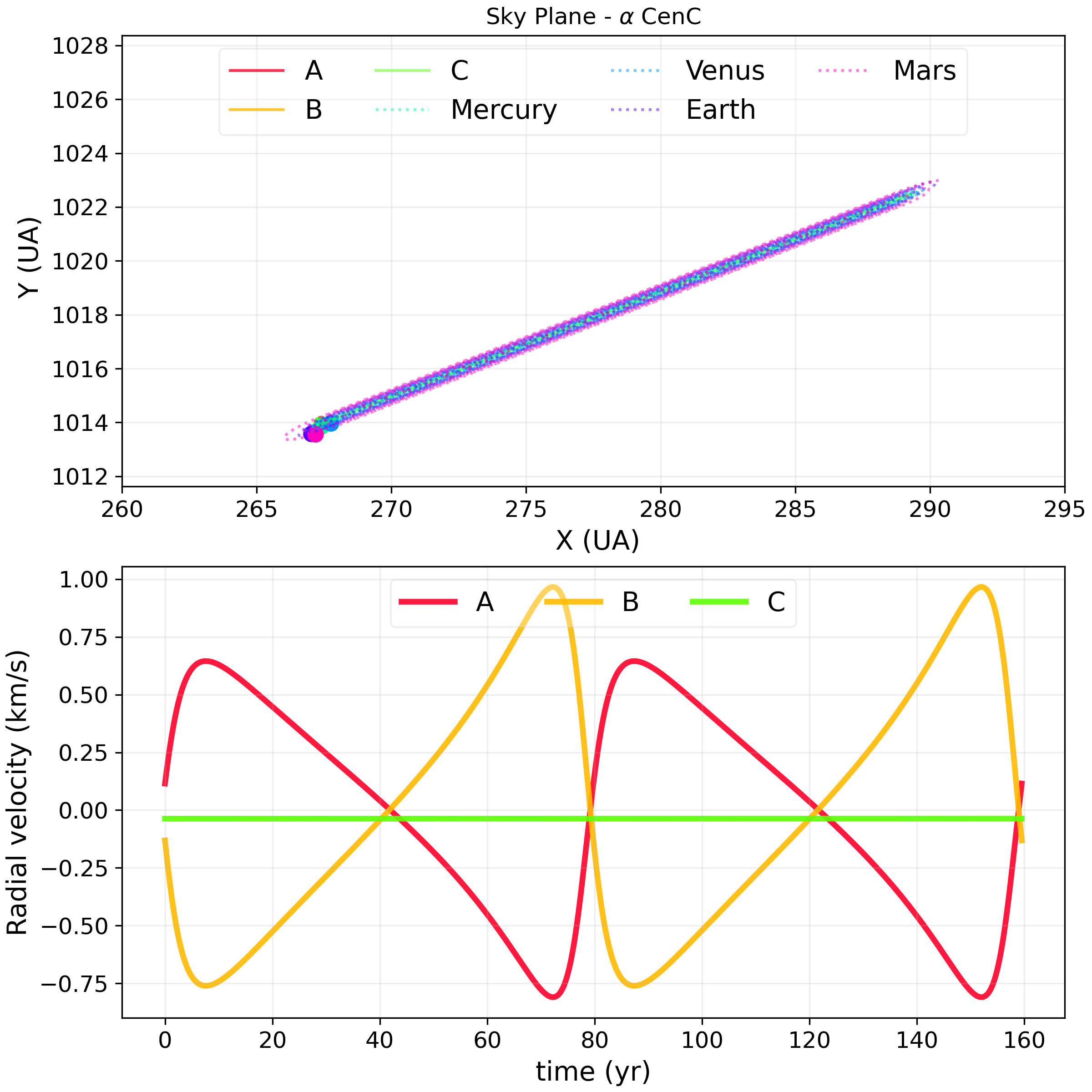}
\end{adjustwidth}
\caption{Each column depicts the evolution of the system with its planets over two periods of the binary, projected onto the plane of the sky (\textbf{upper panels}), and~the estimated radial velocity of the system's three stars (\textbf{lower panels}) In the upper panels, the planets are represented by coloured dots and Alpha Cen A and B by star symbols.}
\label{fig:orbit-RV-curve}
\end{figure}

Figure~\ref{fig:orbit-RV-curve} summarises the results obtained from simulating the triple system with {\tt REBOUND} over two periods of the inner binary, where each star has the four planets of the inner Solar System on S-type orbits. In~the upper panels of the figure, we show the projection of each system on the plane of the sky. These projections were made using the same angles as in Section~\ref{sec:bin_setup}. In~the lower panels, we plot the radial component of the velocity for each star (from left to right: A, B, and C). At~first glance, there appear to be no significant deviations in the radial velocity signals due to the presence of~planets.

\begin{figure}[H]
\includegraphics[width=\columnwidth]{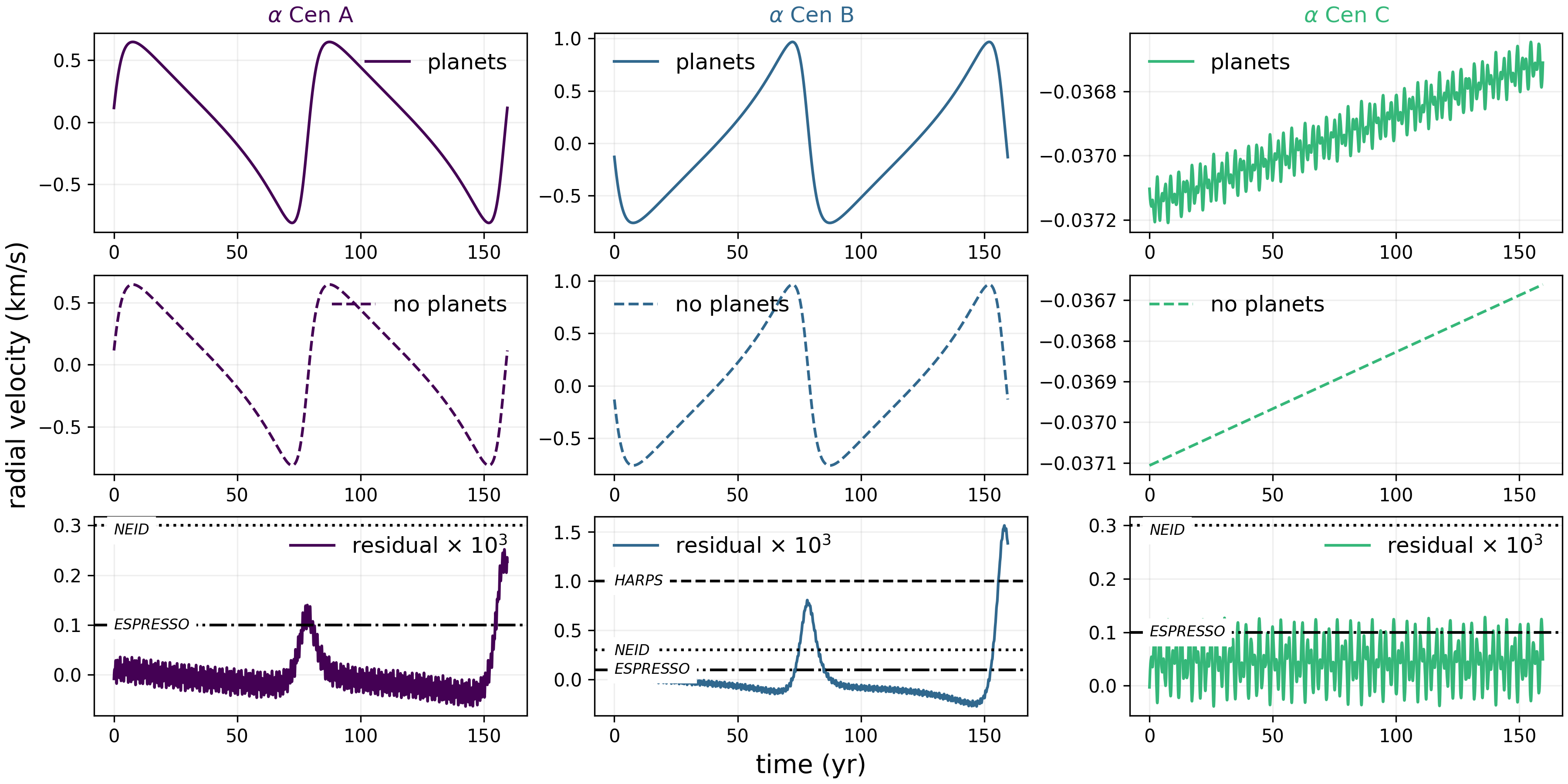}
\caption{Temporal evolution of the radial velocity for each star in the Alpha Cen triple stellar system: A (\textbf{left}), B (\textbf{centre}), and C (\textbf{right}). The~upper panels display the system's evolution with orbiting planets. The~middle panel presents the star's radial velocity in the absence of planets, while the lower panel shows the result of subtracting the two aforementioned signals. The~lowermost panels also illustrate the minimum radial velocity resolutions achievable by the HARPS, NEID, and~ESPRESSO spectrometers.}
\label{fig:RV-curve}
\end{figure}

We therefore conducted an additional simulation, which was similar to the previous integration but without including any planets. This allowed us to subtract one signal to the other, for~each star, in~order to reveal the planetary features. Figure~\ref{fig:RV-curve} presents the outcome of these simulations, where the bottom row shows the residual signal in the radial velocity curves. In~addition, we plotted the lower limits in resolution corresponding to NEID and ESPRESSO spectrometers. The~corresponding value for HARPS is not shown, as its value of 1 m/s is outside the scale presented in these graphs. From~this analysis, we infer that S-type configurations analogous to the inner Solar System could in principle be detected (around A, B, and~C) using~NEID and ESPRESSO. We note however that the radial velocity signals are remarkably weak, with~their semiamplitudes being mostly below 1 m/s, which renders their detection feasible but extremely~challenging.

\unskip

\subsection{Astrometry}
\label{sec:astrometry}

\begin{figure}[H]
\includegraphics[width=\columnwidth]{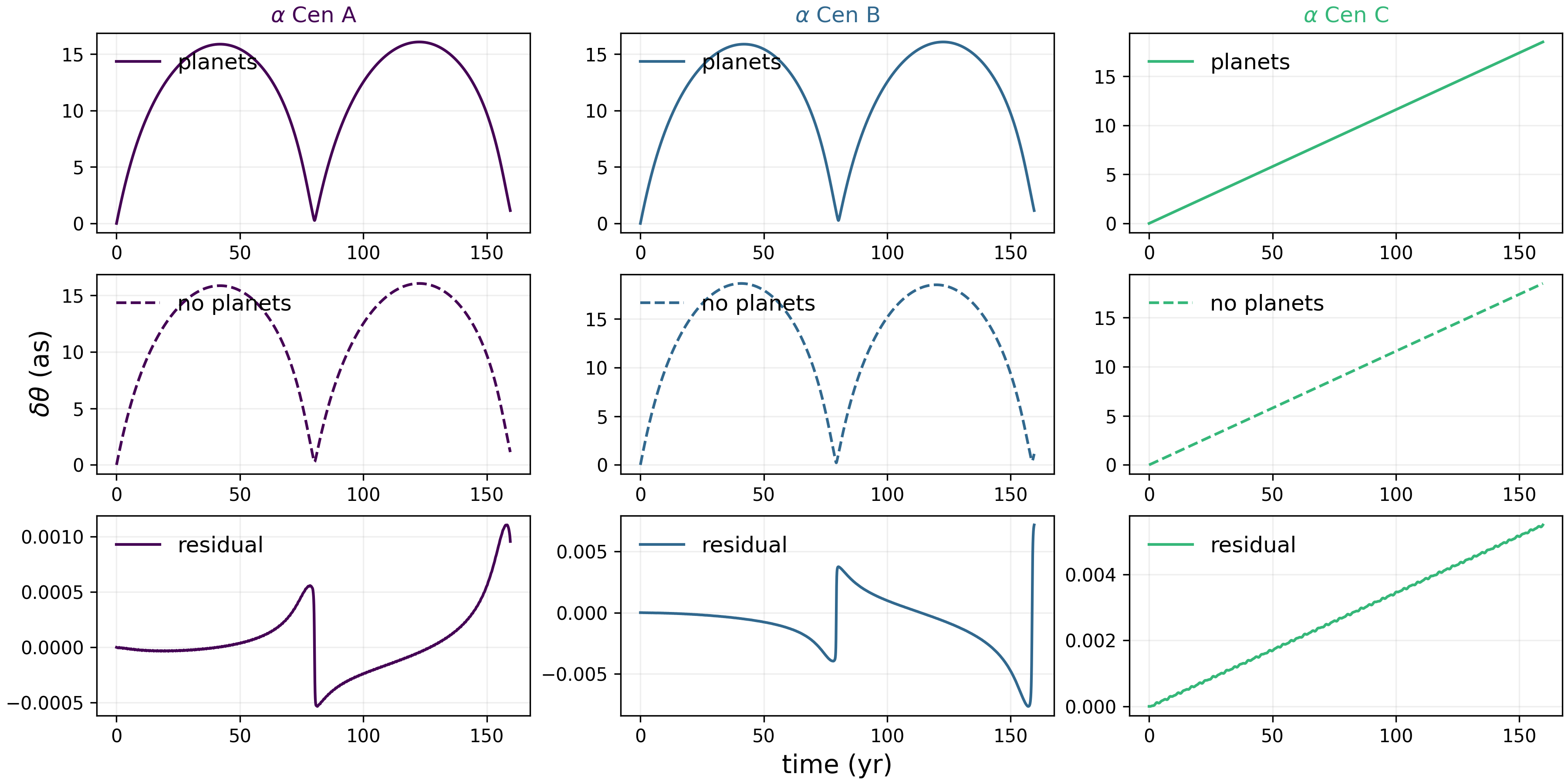}
\caption{\ms{Temporal evolution of the radial velocity for each star in the Alpha Cen triple stellar system: A (\textbf{left}), B (\textbf{centre}), and C (\textbf{right}). The~upper panels display the system's evolution with orbiting planets. The~middle panel presents the star's radial velocity in the absence of planets, while the lower panel shows the result of subtracting the two aforementioned signals.}}
\label{fig:astrometry-curve}
\end{figure}

\ms{Given the proximity of the Alpha Centauri system, we assessed the potential influence of hypothetical planets on the apparent motion of the stars in the celestial plane. This assessment supplements the radial velocity (RV) measurements discussed earlier. Building on the stability regions identified, we considered a low-mass, rocky planetary system akin to the inner Solar System to~determine if its astrometric signals could be detected with current technology. We replicated the methodology from the previous section, thus simulating the stellar system displacement from the initial position $\delta r = r(t) - r_\mathrm{0}$ with and without planets over time. In~Figure~\ref{fig:astrometry-curve}, we show the astrometric deviations caused by the mutual gravitational perturbations between stars and planets around Alpha Cen (upper row)---as opposed to a configuration without any planets (middle row). We found that the deviations (lower row) due to the presence of the other stars were around \mbox{17 arcsecond}s (as) over a binary period, whereas the astrometric deviations due to the presence of planets were much more subtle, thereby ranging between milliarcseconds (mas) and tens of mas.}

\ms{It is worth noting that Gaia's astrometric measurement precision enables the detection of these subtle deviations. As~a matter of fact, Gaia boasts a sensitivity of up to 7 microarcseconds ($\mu$as) for bright stars, thus implying that the milliarcsecond-scale\endnote{\url{https://www.esa.int/Science_Exploration/Space_Science/Gaia/Gaia_factsheet} (accessed on 17 January 2024)} deviations induced by rocky planets in Alpha Centauri are within its detection range. Gaia's sensitivity not only corroborates our estimates but also paves the way for future observations that could directly confirm the existence of such planetary systems.}

\unskip   

\section{Discussion and~Conclusions}
\label{sec:discussion}

Through the present work, we sought to contribute to the growing body of knowledge surrounding the Alpha Centauri system. Below, we discuss some relevant aspects connected to the results presented~here.

\subsection{Disc Evolution and Planet Formation in Alpha~Centauri}

The way discs evolve in binaries is key to understand the distribution of available material to form planets. In~particular, circumstellar discs are heavily truncated by the periodic perturbations of the outer stellar companion. In~the case studied here ($a_{\rm b}=23.3$~au, $e_{\rm b}=0.51947$), the~discs were 10~au in size initially and ended up with radii of about 3~au after tidal truncation. This corresponds to about 12\% of the binary semimajor axis, given that the binary is eccentric. As~a reminder, for~circular binaries, the~discs are truncated between 33\% and 50\% $a_{\rm b}$ \citep{Pichardo2005}. Regardless of the specific values of the disc masses, a~circumstellar disc in a binary exhibits a steeper surface density profile compared to a disc around a single star. This effect unavoidably accelerates the radial inward drift of dust towards the central star \citep{Zagaria2021}. In~other words, there is less time available to transform dust into planetesimals. However, we also note that since more material is concentrated in a more compact disc, this naturally leads to higher values of surface density. This could help to reach higher values of the dust-to-gas ratio within the inner disc regions, which may fuel streaming, self-induced, and/or magneto-hydrodynamical instabilities \citep{Lesur2022}, for~a~review.

Interestingly, at~the onset of disc formation, it is possible to exchange material from one disc to the other as tidal truncation operates. This implies that some material from the outer regions of the disc around B may land onto the disc around A (and vice~versa). The~fact that this alien material was well distributed within both discs (see Figure~\ref{fig:alien-gas}) means that solid bodies with different compositions (e.g., ices and silicates) are more efficiently mixed in binary systems---as opposed to discs around single stars. This could potentially accelerate the growth of dust particles within the disc, but~this statement remains highly speculative at this stage. The~detailed study of dust dynamics in the circumstellar discs of Alpha Cen is left for future~work.

At any rate, in~order to efficiently transform planetesimals into planets within discs, the~eccentricity oscillations of the disc must remain under a certain threshold. This is key to ensure that solid bodies do not collide at destructive relative speeds \citep{Marzari2000, Thebault2009, Marzari2019}. The~detailed modelling of the disc around Alpha Cen B considering a broad variety of disc \mbox{parameters \citep{Martin2020}} revealed that two eccentricity oscillation regimes are expected to occur. However, the~mean eccentricity of the discs eventually reached a quasisteady value around 0.05--0.1, with~little or no disc precession at all (depending on disc viscosity). Therefore, although~the discs in Alpha Cen are part of a multiple stellar system, their eccentricities were shown to remain fairly low, which should have a limited impact on planet formation within the gaseous disc. However, as~the disc evolves, the~gas is progressively removed from the disc, and dust particles decouple from the gaseous disc\endnote{At this stage, an~N-body treatment is more appropriate as opposed to the hydrodynamical approach.}. When this occurs, the~average disc eccentricity increases, thereby rendering collisions between dust particles and solid bodies more destructive~\cite{Martin2020}. So, the~most favourable scenario to form planets in Alpha Cen is during the early gas-rich phase and at stellocentric distances below 3~au. Given the potential mass reservoir of several tens of Earth masses, this means that one or two S-type multiplanetary systems (of rocky planets) may have formed in Alpha Cen~AB.

\subsection{Possible Orbits and Planetary Architectures in Alpha~Centauri}

Given its high eccentricity, the~Alpha Cen AB binary system constitutes a potentially hostile environment for the existence of S-type planets. The~close stellar interactions, with~distances of minimum approach of about 12~au, are likely to have had a significant influence on planetary formation, thus reducing the available mass for such processes and imposing chaotic effects on the surviving planets during each periastron approach. These factors are the primary reasons why the search for exoplanets in this system has been predominantly oriented towards rocky planets, as~opposed to Jovian types, which would be significantly more unstable. In~Section~\ref{sec:disc_dyn}, we presented findings showing that the {\tt MEGNO} chaos indicator, when correlated with the final mass and extent of the expected disc, revealed a notable trend: regions in closer proximity to the stars exhibited increased stability for rocky-type planets. This relationship between the {\tt MEGNO} indicator and the characteristics of the disc provides valuable insights into the dynamics of planetary stability in Alpha~Cen.

Additionally, it is important to acknowledge that while an analysis for 10,000 years does not guarantee dynamical stability during several Gyr, it remains a useful guide for the exoplanetary search in multiple stellar systems. For~example, it is possible to identify stable regions on Gyr scales using inherently more computationally intensive \mbox{integrations \citep{Quarles2016}}. \ms{Nevertheless, to~ensure that the identified regions truly correspond to stability zones, we extended the integration time to 100 kyr. This required a reduction in spatial resolution, thereby confining our analysis to a 20  $\times$  20 grid within the same $a$--$e$ parameter space, which is detailed in Apendix~\ref{sec:appendix_stab_longterm}. This approach balances the need for longer-term stability assessments with practical constraints on computational resources.} Based on this approach, recent studies suggest that up to nine planets (five around A and four around B) could survive on nested prograde orbits for the current age of Alpha Cen AB \citep{Quarles2018}. This kind of methodology was also used in the context of the restricted three-body problem to explore the stability of Earth-like planets within~the habitable zones of Alpha Centauri's stars A and B \citep{Andrade-Ines2014, Giuppone2017}. These previous findings showed that rocky planets are capable of maintaining stable orbits, despite potential high eccentricities and especially at inclinations under 40 degrees. Our proposed method here is complementary, since it enables us to robustly identify stability and chaos regions when considering different orbital parameters within multiple stellar systems. The~main advantage of {\tt MEGNO} over other methods is that the cost of the integration is relatively modest. The~implementation of {\tt MEGNO} maps in the context of exoplanet surveys is relevant with the specific goal of narrowing down the parameter space of possible orbital and physical~parameters.

\subsection{The Search for (Habitable) Exoplanets in Alpha Centauri}

The discovery of exoplanets in the Alpha Cen system could revolutionise our understanding of planet formation and characterisation. In~this context, we highlight the meticulous search carried out by Zhao~et~al. \citep{Zhao2018} combining 10 years of radial velocity measurements (taken with the ES, HARPS, CHIRON, and UVES instruments) with N-body numerical simulations in order to discern hypothetical planetary signals from the observational noise. This study identified the detection thresholds for planets in the classically defined habitable zones for Alpha Cen A and B, as well as~Proxima Centauri. The~possible mass range for stable and detectable planets is up to several tens of Earth masses, which is in~agreement with our analysis. According to the {\tt MEGNO} indicator, we found that rocky planets could cover a wider range of semimajor axes and eccentricity values while remaining stable. Regarding planet detectability, we showcased the capabilities of state-of-the-art spectrographs at the highest resolution possible. Our results indicate that the detection of these small rocky planets in Alpha Cen is feasible in principle, with~campaigns similar to \citep{Zhao2018} but~using the maximum radial velocity resolutions allowed by the NEID and ESPRESSO~spectrometers.

Given its proximity to Earth, this triple stellar system represents the most feasible destination for future interstellar exploration missions, thereby offering a unique opportunity to study a planetary system in the solar vicinity. Various research initiatives, such as the Breakthrough Starshot project\endnote{Breakthrough Starshot Initiative: \url{https://breakthroughinitiatives.org} (accessed on 17 January 2024)}, aim to explore Alpha Centauri in greater detail. This ambitious endeavour plans to send miniature spacecraft to the system at approximately 20\% the speed of light, thereby enabling them to arrive in just over two decades. Moreover, as~a binary system, Alpha Centauri provides a natural laboratory to investigate the formation and stability of planets in habitable zones under complex gravitational and radiative influences. Previous works have already focused on the detection of planets within the circumstellar habitable zones of Alpha Cen \citep{Wang2022} and references therein. Our results are in agreement with previous findings and indicate that Alpha Cen AB could host rocky planets similar to the ones in the inner Solar System. From~an astrobiological perspective, finding such exoplanets would be highly relevant given the remarkable similarities in terms of physical characteristics of both stars to our~Sun.

Regarding Proxima Centauri, given its large distance from AB, we find that planets are highly stable around this red dwarf. Therefore, an~S-type planetary system with habitable planets could thrive around it, e.g., see~\citep{Livesey+2024}. However, despite its potential for habitability and the recent detection of an exoplanet \citep{Anglada2016}, Proxima Centauri is known for frequent stellar flares that could pose significant challenges to the emergence of life \citep{MacGregor2021}. These flares release vast amounts of energy, thereby potentially capable of stripping away the atmospheres of nearby planets---thus heavily damaging hypothetical life forms similar to the ones found on Earth. At~any rate, finding habitable planets around neighbour stars (like Alpha Centauri) would be crucial for humanity, as they would constitute a potential new homes beyond the Solar System. Therefore, the~exploration of the Alpha Cen system, besides~being relevant for the field of planet formation, is also important to secure our distant future in the~cosmos.

\vspace{6pt} 

\authorcontributions{The~authors contributed equally to this work in terms of conceptualisation, methodology, software, validation, and~writing. All authors have read and agreed to the published version of the manuscript.}

\funding{This research was funded by the European Research Council (ERC) under the European Union Horizon Europe research and innovation program (grant agreement No. 101042275, project Stellar-MADE). M.S. acknowledges support from the ANID (Agencia Nacional de Investigación y Desarrollo) through the FONDECYT postdoctoral 3210605. M.S. thanks the ANID---Millennium Science Initiative Program---NCN19\_171. This research has made use of NASA's Astrophysics Data System bibliographic services. The~Geryon cluster at the Centro de Astro-Ingenieria UC was extensively used for the calculations performed in this paper. BASAL CATA PFB-06, the~Anillo ACT-86, FONDEQUIP AIC-57, and~QUIMAL 130008 provided funding for several improvements to the Geryon cluster.}

\dataavailability{The data presented in this study and the details about the software resources supporting the reported results are openly available in GitHub at \url{https://nicolascuello.github.io/Stellar-MADE/alphacen} (accessed on 17 January 2024).}

\acknowledgments{We acknowledge the intensive discussions during ``The Alpha Centauri system: Towards new worlds'' workshop, which was held in Nice from 26 June--30 June 2023 and organised by Lionel Bigot and collaborators. We also thank Martina Toscani, Mart\'in Sucerquia, Valentina Cuello, Mat\'ias Montesinos, and~\'Alvaro Ribas for their kind support and useful suggestions.}

\conflictsofinterest{The authors declare no conflicts of~interest.}

\appendixtitles{yes} 
\appendixstart
\appendix
\section[\appendixname~\thesection]{Planet Stability around Alpha Cen C}
\label{sec:appendix_stab}

\begin{figure}[H]
\begin{adjustwidth}{-\extralength}{0cm}
\centering 

\includegraphics[width=0.43\columnwidth]{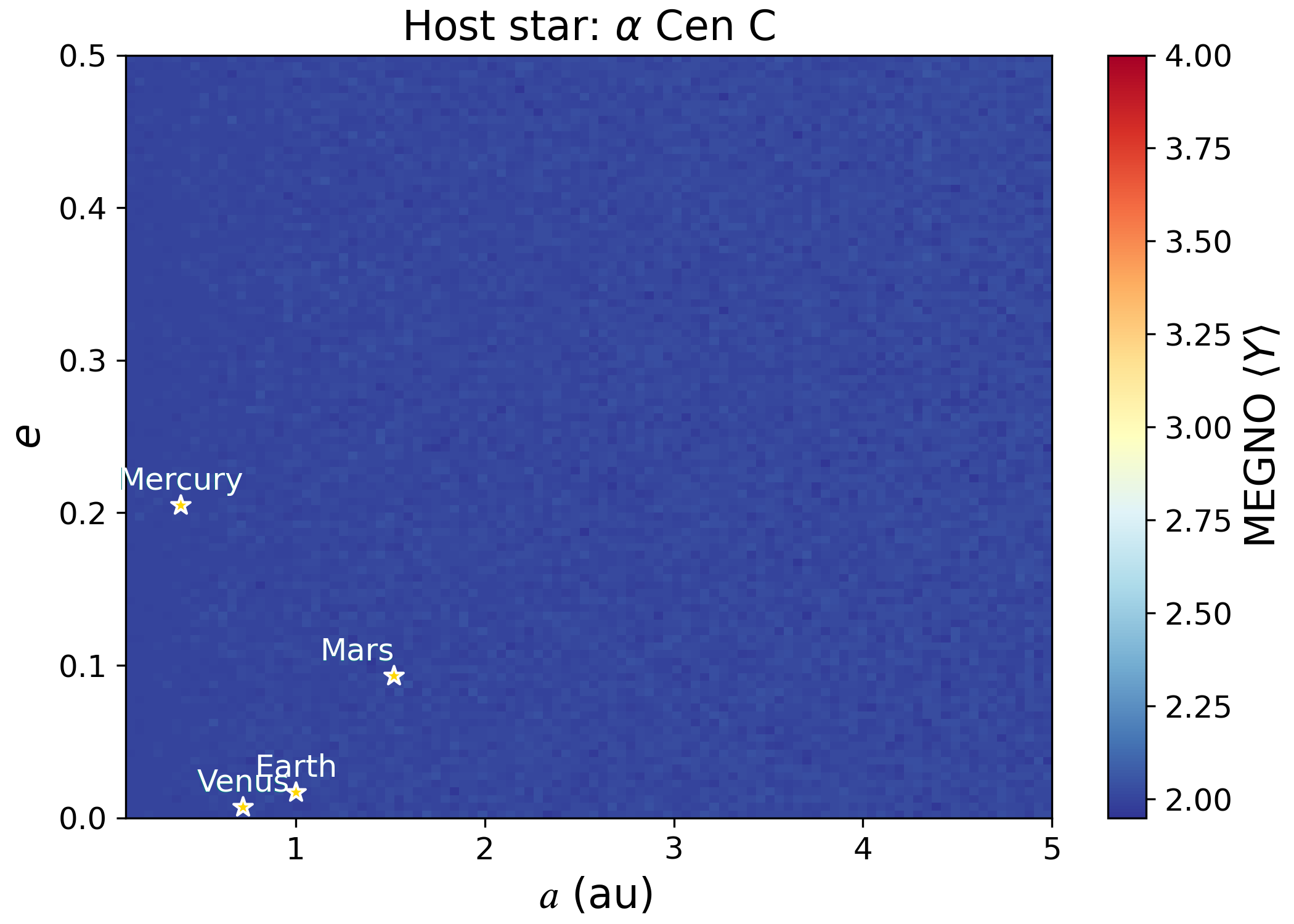}
\includegraphics[width=0.43\columnwidth]{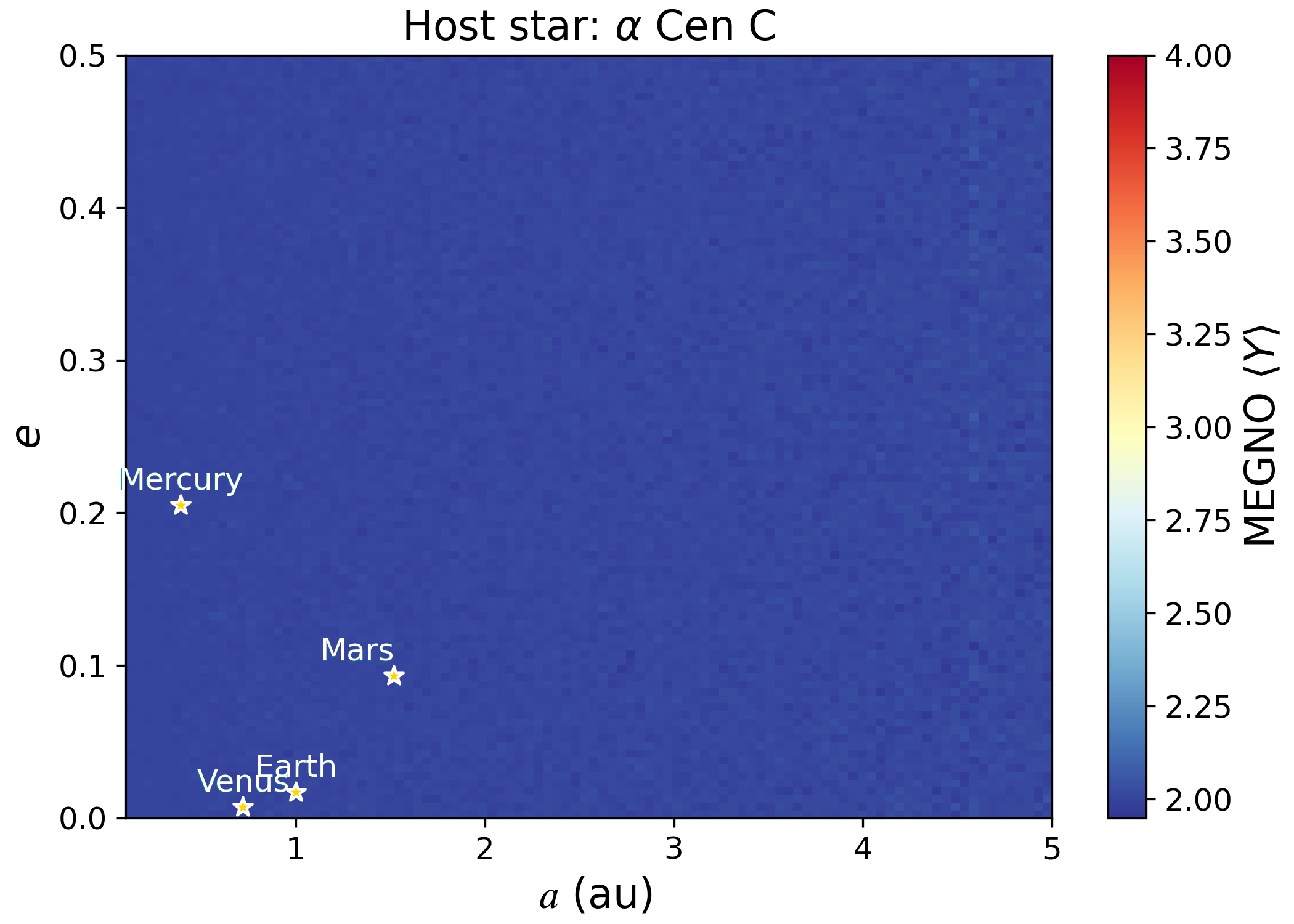}
\includegraphics[width=0.43\columnwidth]{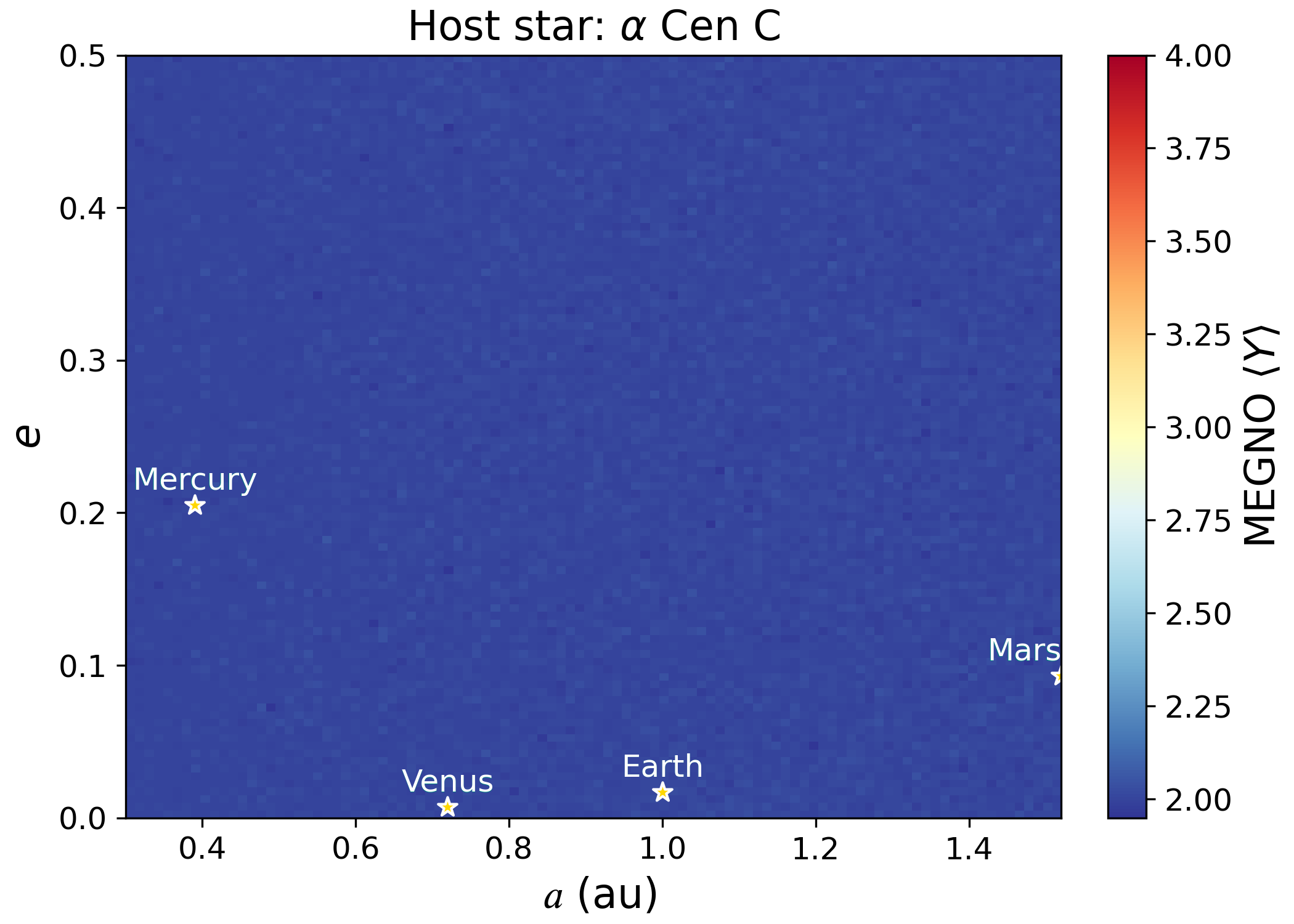}
\end{adjustwidth}
\caption{Stability maps based on {\tt MEGNO} values for test particles (\textbf{leftmost panel}), one Earth-mass planet (\textbf{middle panel}), and~inner Solar System analogues (\textbf{rightmost panel}) over~a 10,000-year period around Proxima Centauri (Alpha Cen C). Dynamically stable regions are coloured in~blue. \label{fig:megnom0XXX}}
\end{figure}

Figure~\ref{fig:megnom0XXX} shows the {\tt MEGNO} stability maps for massless planets---one Earth-mass planet and~inner Solar System analogues---around Proxima Centauri. Given that this red dwarf is at several thousand of au from the inner binary AB, the~planets remain practically unperturbed by the other stars. Therefore, S-type planetary architectures could in principle thrive around Proxima Centauri and maintain stable orbits for long~periods.

\unskip

\section[\appendixname~\thesection]{Integrations Over Longer Time Scales}
\label{sec:appendix_stab_longterm}

\begin{figure}[H]
\includegraphics[width=0.48\columnwidth]{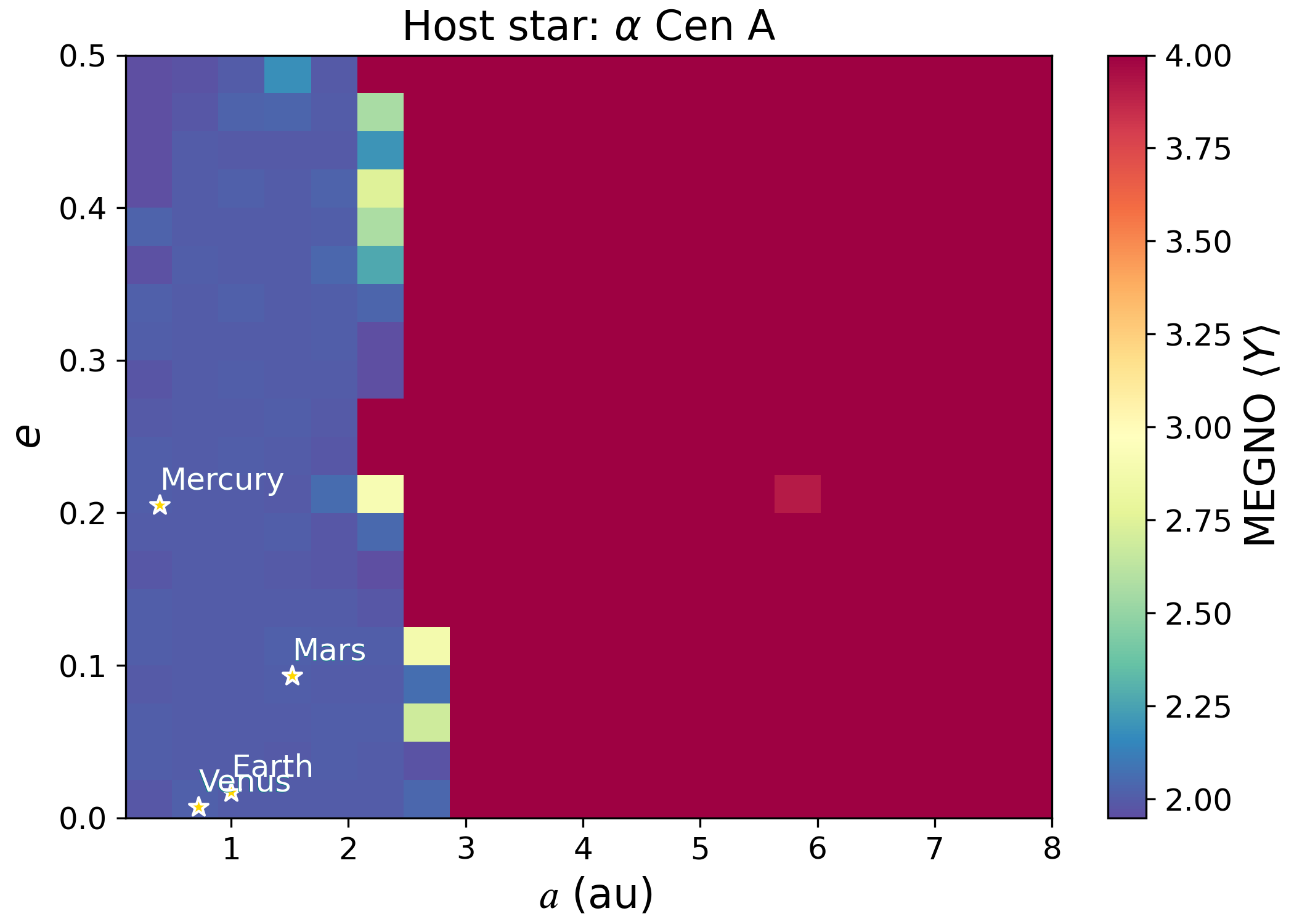}
\includegraphics[width=0.48\columnwidth]{figures/dynamics/MEGNO_Map_HostStar_A_100000yr_0earthmass.png}
\caption{Stability maps based on {\tt MEGNO} values for test particles around Alpha Cen A (\textbf{leftmost panel}) and Alpha Cen B (\textbf{leftmost panel}) over a 100,000-year period for a 20  $\times$  20 grid in the $a$--$e$ parameter space. Dynamically stable regions are coloured in~blue. \label{fig:megno_t1e5}}
\end{figure}

\ms{Figure~\ref{fig:megno_t1e5} shows the {\tt MEGNO} stability maps for massless planets around Proxima Centauri A and B spanning a timescale of up to 100,000 years. These maps feature a lower spatial resolution (20  $\times$  20) as~opposed to the ones in Section~\ref{sec:stability} (100  $\times$  100 resolution). The~goal of this test was to verify the stability of the regions identified in the earlier high-resolution figures. Furthermore, in~these simulations, the~spatial domain has been extended to eight astronomical units, thus seeking potential stability zones in the outermost areas of the systems. Broadly speaking, the~regions classified as stable in Figure~\ref{fig:megnom0} coincide with those found in this new set of integrations, wherein test particles have also been utilised. This justifies our proposed approach.}

\begin{adjustwidth}{-\extralength}{0cm}

 \printendnotes[custom] 
\reftitle{References}

\PublishersNote{}
\end{adjustwidth}
\end{document}